\documentclass[twocolumn]{aastex631}

\renewcommand{\ion}[1]{{\normalfont~\textsc{#1}}}

\usepackage{longtable}
\usepackage{textcomp}
\usepackage{graphicx}	
\usepackage{amssymb,amsmath} 
\usepackage{hyperref}
\usepackage{booktabs}
\usepackage{gensymb}
\usepackage{listings}
\usepackage{lipsum}
\usepackage{float}
\usepackage{tabularx}
\newcolumntype{Y}{>{\centering\arraybackslash}X}
\usepackage{xcolor}
\usepackage{soul}
\usepackage{silence}
\begin{document}

\title{The CIViL$^{\star}$ Survey: The Discovery of a C\ion{iv} Dichotomy in the CGM of L$^{\star}$ Galaxies}

\correspondingauthor{Samantha L. Garza}
\email{samgarza@uw.edu}

\author[0000-0003-4521-2421]{Samantha L. Garza}
\affiliation{Department of Astronomy, University of Washington, Seattle, WA, 98195}

\author[0000-0002-0355-0134]{Jessica K. Werk}
\affiliation{Department of Astronomy, University of Washington, Seattle, WA, 98195}

\author[0000-0002-2606-5078]{Trystyn A. M. Berg}
\affiliation{National Research Council Herzberg Astronomy and Astrophysics, 5071 West Saanich Road, Victoria, B.C., V8Z6M7, Canada}

\author[0000-0003-3520-6503]{Yakov Faerman}
\affiliation{Department of Astronomy, University of Washington, Seattle, WA, 98195}

\author[0000-0002-3391-2116]{Benjamin D. Oppenheimer}
\affiliation{University of Colorado, 
Center for Astrophysics and Space Astronomy, 389 UCB, 
Boulder, CO 80309, USA}

\author[0000-0002-3120-7173]{Rongmon Bordoloi}
\affiliation{Department of Physics, North Carolina State University, Raleigh, NC 27695, USA}

\author[0000-0002-1768-1899]{Sara L. Ellison}
\affiliation{Department of Physics and Astronomy, University of Victoria, Victoria, British Columbia, V8P 1A1, Canada.}



\begin{abstract}



This paper investigates C\ion{iv} absorption in the circumgalactic medium (CGM) of L$^{\star}$ galaxies and its relationship with galaxy star formation rates. We present new observations from the C\ion{iv} in L$^{\star}$ survey (CIViL$^{\star}$; PID$\#$17076) using the Hubble Space Telescope/Cosmic Origins Spectrograph. By combining these measurements with archival C\ion{iv} data (46 observations total), we estimate detection fractions for star-forming (sSFR $>$ 10$^{-11}$ yr$^{-1}$) and passive galaxies (sSFR $\leq$ 10$^{-11}$ yr$^{-1}$) to be 72$_{-18}^{+14}$\% [21/29] and 23$_{-15}^{+27}$\% [3/13], respectively. This indicates a significant dichotomy in C\ion{iv} presence between L$^{\star}$ star-forming and passive galaxies, with over 99\% confidence. This finding aligns with \cite{Tumlinson_2011}, which noted a similar dichotomy in O\ion{vi} absorption. Our results imply a substantial carbon reservoir in the CGM of L$\star$ galaxies, suggesting a minimum carbon mass of $\gtrsim$ 3.03 $\times$ 10$^{6}$ M$_{\odot}$ out to 120 kpc. Together, these findings highlight a strong connection between star formation in galaxies and the state of their CGM, providing insight into the mechanisms governing galaxy evolution.

\end{abstract}



\section{Introduction} 

The diffuse gaseous atmosphere surrounding the star-filled inner region of a galaxy is known as its CGM. The CGM plays an essential role in a galaxy's evolution by hosting the gaseous reservoir that feeds the galaxy with gas, replenishing fuel for star formation, and keeping a record of metal-enriched material ejected from the disk through winds and other feedback processes \citep{Lehner_howk_2011, Peeples_2014, Werk_2014, tumlinson_2017}. The Hubble Space Telescope and the Cosmic Origins Spectrograph (\textit{HST}/COS) have significantly advanced the study of the CGM over the past fifteen years, creating a compendium of gas around dwarf to L$^{\star}$ galaxies \citep[e.g.,][]{werk_2013, Bordoloi_2014, Borthakur_2015, Heckman_2017, berg_2018, Lehner_2018}, with additional focus on specific demographics such as starbursts \citep{Heckman_2017} and active galactic nuclei (AGN) \citep{berg_2018}. However, many questions remain about how ongoing processes within galaxies impact the physical state of their CGM and vice-versa. Quantifying the connection between these halos and their host galaxies can provide essential constraints on the co-evolution of galaxies and the CGM. 

Some progress has been made for the cool phase of the CGM (T $\sim$ 10$^{4}$ K), where metal-enriched gas traced by Mg\ion{ii} has been used to compare the CGM content between star-forming and passive galaxies. The consensus is that the CGM of star-forming galaxies has higher Mg\ion{ii} equivalent widths and covering fractions than that of passive galaxies \citep{Lan_2020, Anand_2021}. Mg\ion{ii} is often associated with star-formation-driven winds \citep{Rubin_2014}, showing a strong incidence rate along the minor axes of star-forming galaxies \citep{Bordoloi_2011}. Other cool gas phase tracers such as H\ion{i}, Si\ion{ii}, and C\ion{iii} show no statistically significant correlation with galaxy star-forming properties \citep{Tumlinson_2013, werk_2013}.  


For the warm phase, a seminal result from the COS-Halos survey found a dichotomy in the content of O\ion{vi} in the CGM of star-forming and passive galaxies \citep{Tumlinson_2011}. O\ion{vi}, a likely tracer of highly ionized T $\sim$ 10$^{5.5}$ K gas, is abundant in the halos of actively star-forming L$^{\star}$ galaxies, while it is rarely found within 150 kpc of L$^{\star}$ passive galaxies \citep{Tumlinson_2011, Johnson_2015, Zahedy_2019}. This dichotomy holds with greater than 3$\sigma$ significance for stellar mass-controlled samples of low-redshift (z $<$ 0.6) galaxies with 10$^{10.1}$ $<$ M$_{\star}$ $<$ 10$^{10.9}$ \citep{Tchernyshyov_2023}.  



While high-ions and low-ions likely trace distinct gas phases of the CGM and show different correlations with galaxy star-forming properties, it is unclear where intermediate ionization state gas tracers such as Si\ion{iv} and C\ion{iv} fit for L$^{\star}$ galaxies. For sub-L$^{\star}$ galaxies (log M$_{\star}$/M$_{\odot}$ $ \lesssim$ 10), a tentative correlation was detected between C\ion{iv} absorption strength and star formation mirroring the dichotomy seen in O\ion{vi} for L$^{\star}$ galaxies \citep{Bordoloi_2014}. A similar case is probable for L $\approx$ L$^{\star}$ galaxies, where these intermediate ions may trace warm photoionized material similar to low-ions and/or may be produced in part by collisional ionization, either in equilibrium or out of equilibrium, for example, by turbulent mixing layers \citep[e.g.][]{Kwak_shelton_2010}.  
However, due to the wavelength coverage of available gratings on \textit{HST}/COS, C\ion{iv} and Si\ion{iv} are rarely observed by CGM surveys prioritizing O\ion{vi}, such as COS-Halos \citep{werk_2013} and COS-GASS \citep{Borthakur_2015}. For this reason, we initiated the C\ion{iv} in L$^{\star}$ galaxies (CIViL$^{\star}$) Survey to fill this notable gap in previous COS absorption galaxy studies by providing NUV coverage of C\ion{iv} for galaxy with existing O\ion{vi} measurements in the CGM. For this work, we assume a flat-universe $\Lambda$CDM cosmology with $H_{0}$ = 67.8 km s$^{-1}$ Mpc$^{-1}$ and $\Omega_{m}$ = 0.308 \citep{Plank2016}.


\section{Observation and Data Analysis}\label{sec: observation_and_data} 

\begin{table*}[]
\caption{CIViL$^{\star}$ Galaxy Sample Properties and C\ion{iv} Measurements}
\centering
\begin{tabular}{cccccccccc}
\hline
Galaxy               & $z_{\rm gal}$        & $z_{qso}$            & sSFR                   & M$_{\star}$              & R$_{\rm proj}$ &  R$_{\rm proj}$/R$_{\rm 200c}$ &  M$_{\rm 200c}$ & N$_{\rm CIV}$          & ref                  \\
\multicolumn{1}{l}{} & \multicolumn{1}{l}{} & \multicolumn{1}{l}{} & (log$_{10}$ yr$^{-1}$) & (log$_{10}$ M$_{\odot}$) & (kpc) & \multicolumn{1}{l}{} & (log$_{10}$ M$_{\odot}$) & (log$_{10}$ cm$^{-2}$) & \multicolumn{1}{l}{} \\
(1)                  & (2)                  & (3)                  & (4)                    & (5)                      & (6)   & (7)                    & (8) & (9) & (10)                  \\ \hline
J1427+2629\_45940    & 0.0330               & 0.364                & -12.0                    & 10.4                     & 170   & 1.07 & 11.6 & \textless 13.14        & COS-GASS                 \\
J1502+0649\_41743    & 0.0460               & 0.288                & -10.2                  & 10.5                     & 224   &1.29 & 11.8 & \textless 13.55        & COS-GASS                 \\
J1544+2740\_28317    & 0.0320               & 0.163                & \textless -12.1        & 10.1                     & 196   & 1.49 & 11.4 & \textless 12.72        & COS-GASS                 \\
J0226+0015\_268\_22  & 0.2274               & 0.615                & \textless -11.8        & 10.8                     & 80   & 0.32 & 12.3 & \textless 13.13        & COS-Halos                \\
J0401-0540\_67\_24   & 0.2197               & 0.570                & -10.1                  & 10.1                     & 83   & 0.60 & 11.5 & \textgreater 14.26     & COS-Halos                \\
J0950+4831\_177\_27  & 0.2119               & 0.589                & \textless -11.7        & 11.2                     & 92   & 0.16 & 13.4 & \textgreater 14.22     & COS-Halos                \\
J1016+4706\_274\_6   & 0.2520               & 0.822                & -10.4                  & 10.2                     & 23   & 0.16 & 11.6 & \textgreater 14.48     & COS-Halos                \\
J1016+4706\_359\_16  & 0.1661               & 0.822                & -10.4                  & 10.5                     & 44   & 0.26 & 11.8 & \textgreater 14.66     & COS-Halos                \\
J1342-0053\_157\_10  & 0.2270               & 0.326                & -10.2                  & 10.9                       & 35    & 0.11 & 12.6 & \textless 13.71        & COS-Halos                \\
J1342-0053\_77\_10   & 0.2013               & 0.326                & \textless -11.0        & 10.5                     & 32    & 0.19 & 11.8 & \textless 12.87        & COS-Halos                \\
J1419+4207\_132\_30  & 0.1792               & 0.873                & -9.5                   & 10.6                     & 88    & 0.46 & 12.0 & \textgreater 14.52     & COS-Halos                \\ \hline
\end{tabular}
\label{tab: galaxy_properties}
\tablecomments{Comments on columns: (1) Galaxy name: for the COS-Halos galaxies it SDSS field identifier and galaxy identifier, where the first number
is the position angle in degrees from the QSO and the second number is the projected separation in arcseconds (impact parameter) from the QSO, for the COS-GASS the value is the COS-GASS ID; (2) galaxy redshift; (3) QSO redshift (4) specific star formation rate (sSFR): For more details on how the sSFRs were calculated for the galaxies in sample, refer to \citep{werk_2012, Bordoloi_2014, Borthakur_2015, Garza_2024}. On average, for galaxies of these masses, sSFR errors will be on the order of a few - several tenths of a dex.; (5) stellar mass: stellar masses are accurate to about $\sim$ 50\%; (6) impact parameter; (7)impact parameter normalized by virial radius ; (8) virial halo mass (9) C\ion{iv} column density; (10) which survey the galaxy was matched to, either COS-Halos \citep{werk_2013} or COS-GASS \citep{Borthakur_2015}.}
\end{table*}

\subsection{Sample Selection}\label{sec: sample_selection}


The CIViL$^{\star}$ Survey (PID$\#$17076) more than doubles the number of literature sightlines that probe C\ion{iv} in the inner CGM 
of L$^{\star}$ galaxies. It consists of nine UV-bright QSOs, probing the halos of eleven low redshift ($z \lesssim$ 0.25) galaxies. One of the main goals of this survey is to observe C\ion{iv} in galaxies that have published O\ion{vi} detections. To build the survey, we identified galaxies from COS-Halos \citep{werk_2013} and COS-GASS \citep{Borthakur_2015} that would provide a valuable legacy dataset of C\ion{iv} coverage within one virial radius. This sample of eleven galaxies is representative of the broad range in galaxy stellar mass (log$_{10}$M$_{\star}$/M$_{\odot}$ $\sim$ 10.1-11.4) and impact parameter (R$_{\rm proj} \sim$ 20-224 kpc) of the L$^{\star}$ galaxies observed in previous COS-CGM programs \citep[e.g.,][]{werk_2013, Bordoloi_2014, Borthakur_2015, Heckman_2017, berg_2018, Garza_2024} in order to enable fair comparisons to other galaxies types in different stages of evolution. 

In particular, the final sample was selected by maximizing the number of $\sim$L$^{\star}$ galaxies that are simultaneously control matched in both M$_{\star}$ and R$_{\rm proj}$ within $\pm0.2$~dex \citep[see][for more details]{berg_2018} to previous COS-CGM programs that include a variety of special evolutionary phases of galaxies (i.e.~starburst or AGN). Furthermore, the ratio of impact parameter to virial radius of the survey is well matched (within $\pm0.3$~dex) to the distribution in the COS-Dwarfs \citep{Bordoloi_2014} survey which enables a similar comparison of radial profiles over the two decades of galaxy mass in a systemic fashion with C\ion{iv}. At least 3 `normal' galaxies were control matched to the starburst or AGN galaxies, whilst at least 5 were matched to COS-Dwarf galaxies. Work utilizing this controlled match aspect of the sample to uncover the nature of highly ionized gas surrounding AGN hosts is forthcoming in Berg et al.~(in prep.). The focus of this paper is the comparison between passive and star-forming $\sim$L$^{\star}$ galaxies. 


To create an average profile, normalized by the virial radius, we need the virial radii and virial masses. To estimate dark matter halo masses we use the same methodology as \cite{Tchernyshyov2022}. We use the stellar mass-halo relation defined in \citet[][see Tab J1 for fit parameters]{Behrooz_2019} in combination with the technique described in \cite{Hu_2003}. This technique converts the calculated halo masses, estimated using the galaxy stellar mass and redshift, such that the average mass density within the halo radius is 200 times the critical density of the universe. Using our newly converted virial halo masses (M$_{200c}$), we calculate virial radii (R$_{200c}$). These and other galaxy properties can be found in Table \ref{tab: galaxy_properties}.


\subsection{COS Spectroscopy}\label{sec: cos_spec}

The quasar spectra for the CIViL$^{\star}$ Survey were taken using either the G160M  or the G185M grating on the Cosmic Origins Spectrograph \citep[COS;][]{Froning_and_Green_2009, Green_2012} on the Hubble Space Telescope as a part of a 52-orbit Cycle 29 HST Program (PID$\#$17076). The primary spectral features of interest are absorption lines from the C\ion{iv} doublet ($\lambda\lambda$1548, 1550) at the redshift of each host galaxy. The CIViL$^{\star}$ QSOs have FUV magnitudes of $\sim$ 17.0-18.4 and redshifts ranging from 0.032-0.252. Each target QSO was observed between 2-5 orbits in either G185M \citep[targets matched to COS-Halos,][]{werk_2013} or G160M \citep[targets matched to COS-GASS,][]{Borthakur_2015}, depending on the galaxy redshift. Our exposure times were calculated to detect a 50 m\AA~ feature at a confidence of 2$\sigma$. The S/N of the resultant spectra range from 5-12 per resel at the wavelengths of the C\ion{iv} doublet. 

We combine the CALCOS-generated x1D files using v3.1.1 of the COADD\_X1D routine provided by the COS-GTO team \citep{danforth_2016}, which properly treats the error arrays of the input files using Poisson statistics. The IDL COADD\_X1D routine was modified to injest new data from G185M grating. 
The code aligns the different exposures by determining a constant offset determined by cross-correlating strong ISM lines in a 10\AA~wide region of the spectrum. For data taken with the G160M grating, each COS resolution element at R $\sim$18,000 is sampled by six raw pixels at 12.23 m\AA~per pixel. G185M spectra consist of three 35 \AA~ stripes separated by two 64 \AA~ gaps with an R $\sim$18,000 and  typical spectral dispersion of 33 m\AA~per pixel (0.2 \AA~per six-pixel resolution element). Both the G160M and G185M spectra are Nyquist sampled. For G160M we binned to nyquist sampling with 2 bins per resolution element. The spectra for G185M, which are slightly lower resolution, are well sampled and did not require binning. The resulting science-grade spectra are characterized by a FWHM $\approx$ 16 km s$^{-1}$ and $\approx$ 38 km s$^{-1}$ for G160M and G185M data, respectively. We perform continuum fitting with the {\tt linetools} package\footnote{https://github.com/linetools}
, an open-source code for analysis of 1D spectra.


\subsection{Absorption Line Measurements}\label{sec: abs_line_meas}

We determine absorption feature line identifications and redshifts in the CIViL$^{\star}$ spectra using using the {\tt\string PyIGM IGMGuesses} GUI\footnote{https://github.com/pyigm}.
Since the redshifts of the galaxies in the CIViL$^{\star}$ survey are well known (see Table \ref{tab: galaxy_properties}) CGM absorption features were identified by scanning the spectra for features within $\sim$300 km/s associated with the C\ion{iv} doublet ($\lambda\lambda$1548, 1550). If blends or absorption other than C\ion{iv} were present we followed the same identification methodology described in \citet{Garza_2024}. 


\begin{figure*}%
    \centering
    {{\includegraphics[width=0.95\textwidth]{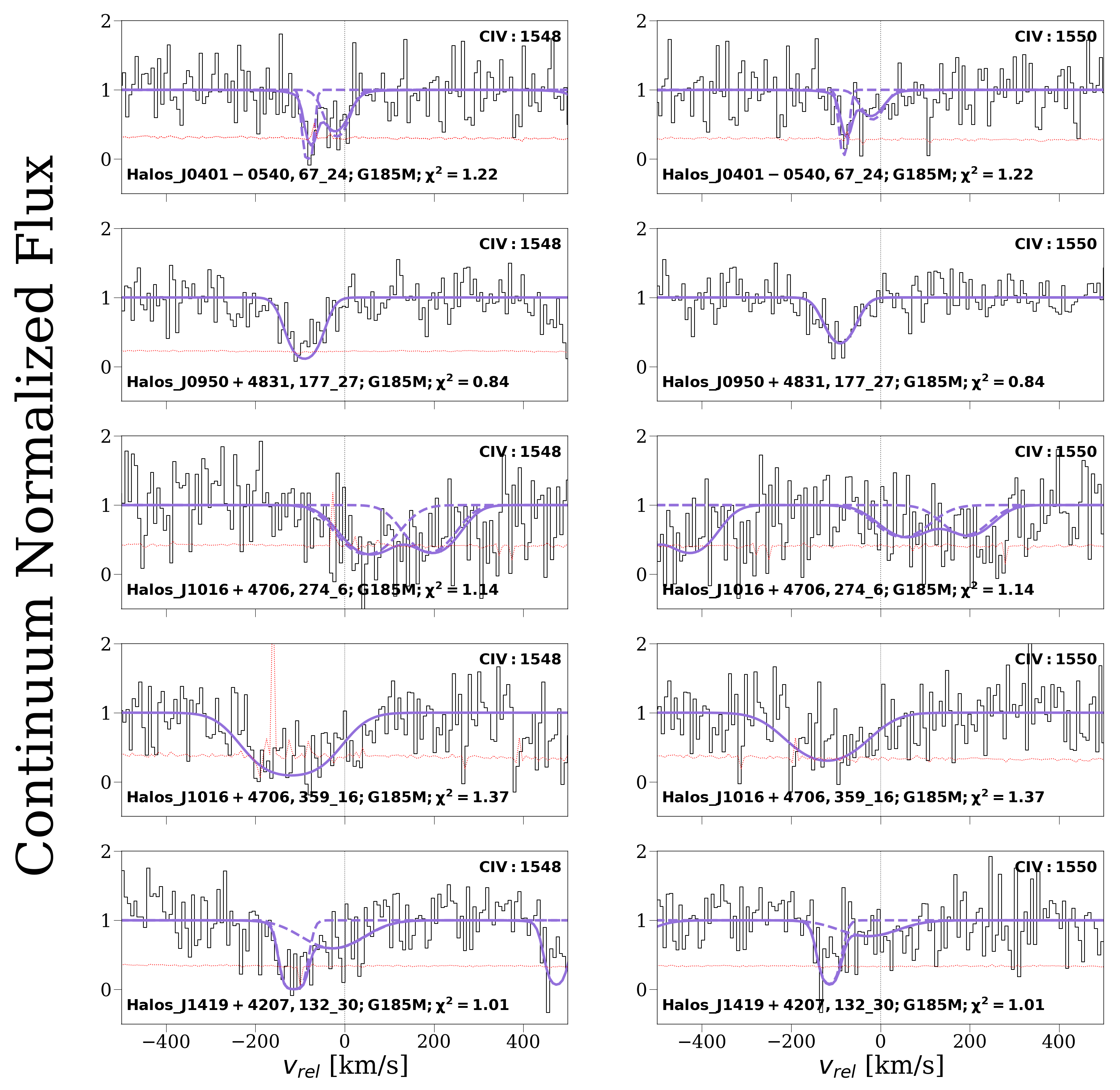}}}%
    \caption{Regions of the HST/COS continuum spectrum showing the C\ion{iv} $\lambda$$\lambda$ 1548 1550 line absorption features of the CIViL$^{\star}$ QSO-galaxy pairs set in the rest frame of each individual galaxy. For lines of sight that have multiple components, the individual fits are shown as the dashed purple line where the sum of the components is shown as the solid purple line. The red line in each spectrum represents the continuum flux error.} The bottom left corner has the QSO-Galaxy IDs and the reduced chi squared fit for the absorption features.%
    \label{fig: spec_abs}
\end{figure*}

Based on the manual identifications from the {\tt\string PyIGM IGMGuesses} GUI, we measure the C\ion{iv} column densities (N$_\mathrm{C IV}$), Doppler parameter ($b$), and the relative velocity of the absorption components ($v_{rel}$) using Voigt profile fitting with the package {\tt\string veeper}\footnote{https://github.com/jnburchett/veeper}.
This package uses {\tt\string scipy.optimize.least\_squares}\footnote{https://github.com/scipy/scipy}
to perform a least squares minimization to obtain its measurements and incorporates the COS line spread function.  Five of our QSO-Galaxy line-of-sight pairs show C\ion{iv} detections (Figure \ref{fig: spec_abs}); however, the absorption features are saturated so we report them as lower limits. When multiple absorption components are found in a galaxy's search window, their column densities are summed, and the resulting total column density is associated with the galaxy. The other five of the line-of-sight pairs show no C\ion{iv} absorption and we report them as upper limits. In these non-detection regions, we calculate a 2$\sigma$ upper limit on the column density as estimated by the apparent optical depth method (AODM) with the {\tt\string linetools XSpectrum1D} package\footnote{https://github.com/linetools/linetools/tree/v0.3}
over a 100 km s$^{-1}$ velocity span centered on the galaxy redshift. By default, we use the stronger line at 1548\AA~ to estimate $2\sigma$ equivalent width upper limits, but in cases where there is blending or contamination, we use the 1550\AA~ line. Our C\ion{iv} column density measurements can be found in Table \ref{tab: galaxy_properties} and they are shown as a function of the impact parameter normalized by virial radius and stellar mass in panels 2 and 3 of Figure~\ref{fig: nciv_rproj_mstar} respectively. 

\subsection{Archival Observations}\label{sec: archival_data}

Previous studies have shown that observable tracers of the CGM depend on galaxy mass, redshift, and environment \citep{Bergeron_1986, Bahcall_1991, Chen_2001, Stocke_2006, Bordoloi_2011, werk_2013, Johnson_2015, Burchett_2016, Tejos_2016, Bordoloi_2018, Tchernyshyov_2023}, making them important variables to consider when studying the differences between the CGM of star-forming and passive galaxies. To address the effect of these galaxy properties, we increase our sample size with CGM C\ion{iv} measurements using published HST/COS data from COS-Halos \citep{werk_2013}, COS-Dwarfs \citep{Bordoloi_2014}, and COS-Holes \citep{Garza_2024}. Eight out of eleven of our QSO-galaxy line-of-sight pairs are COS-Halos \citep{werk_2013} galaxies, thus it is intuitive to include the two galaxies from the COS-Halos Survey that already have C\ion{iv} observations. The COS-Holes Survey\footnote{We note that results from \cite{Garza_2024} suggest that C\ion{iv} absorption does not show obvious variation as a function of SMBH mass. } \citep{Garza_2024} sits well within the stellar mass and impact parameter range of the CIViL$^{\star}$ survey and adds nine observations to the sample. We also include observations from the COS-Dwarfs survey \citep{Bordoloi_2014} because, as mentioned in \S\ref{sec: sample_selection}, the CIViL$^{\star}$ survey is well matched with this survey's ratio of impact parameter to viral radius.  
By including observations from these three surveys we have a total combined C\ion{iv} sample of 65 observations.

\section{Results}\label{sec:Results}




\begin{figure}[h!t]
    \centering
    \includegraphics[width = 0.95\columnwidth]{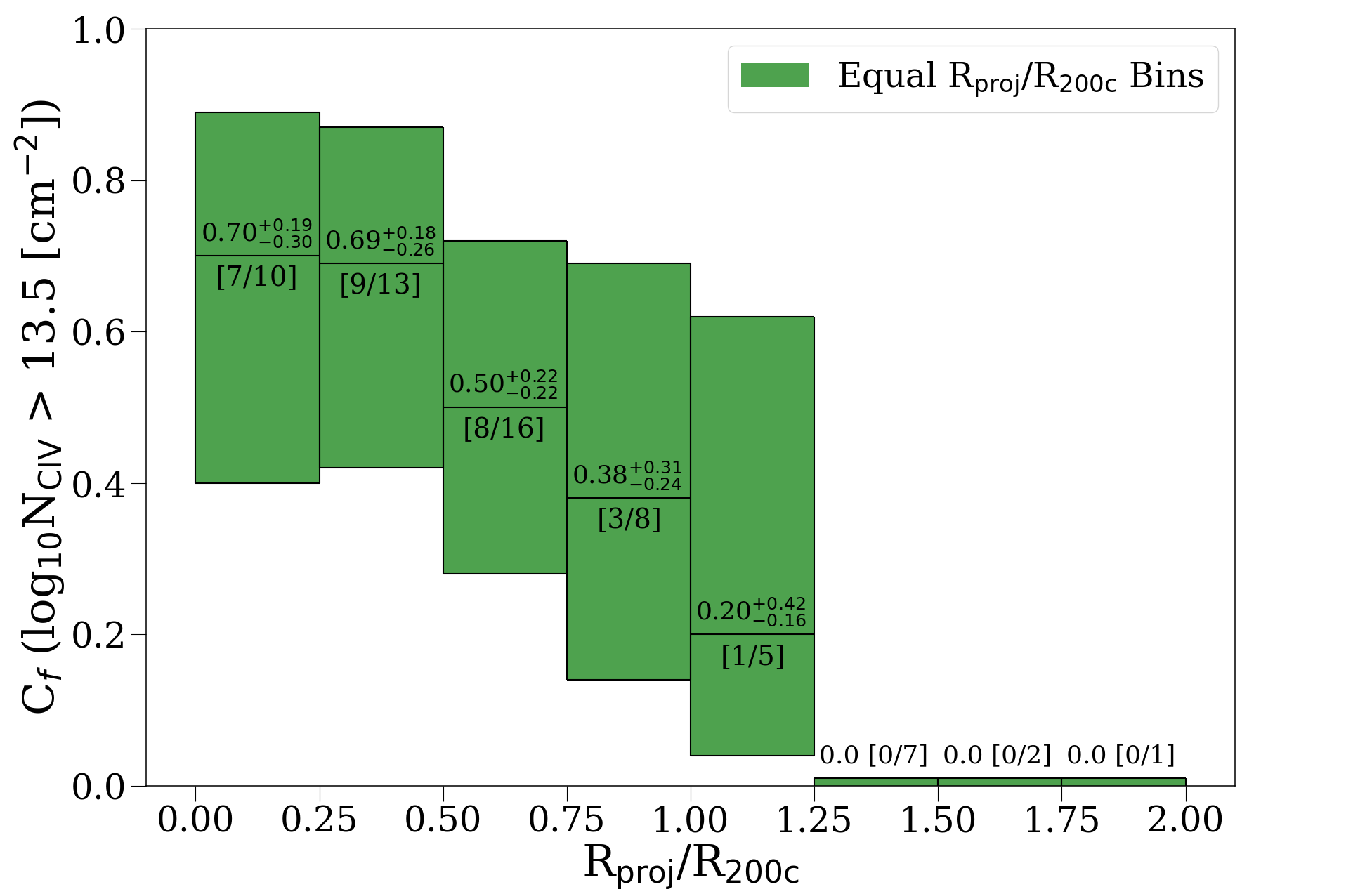}
     \qquad
    \includegraphics[width = 0.96\columnwidth]{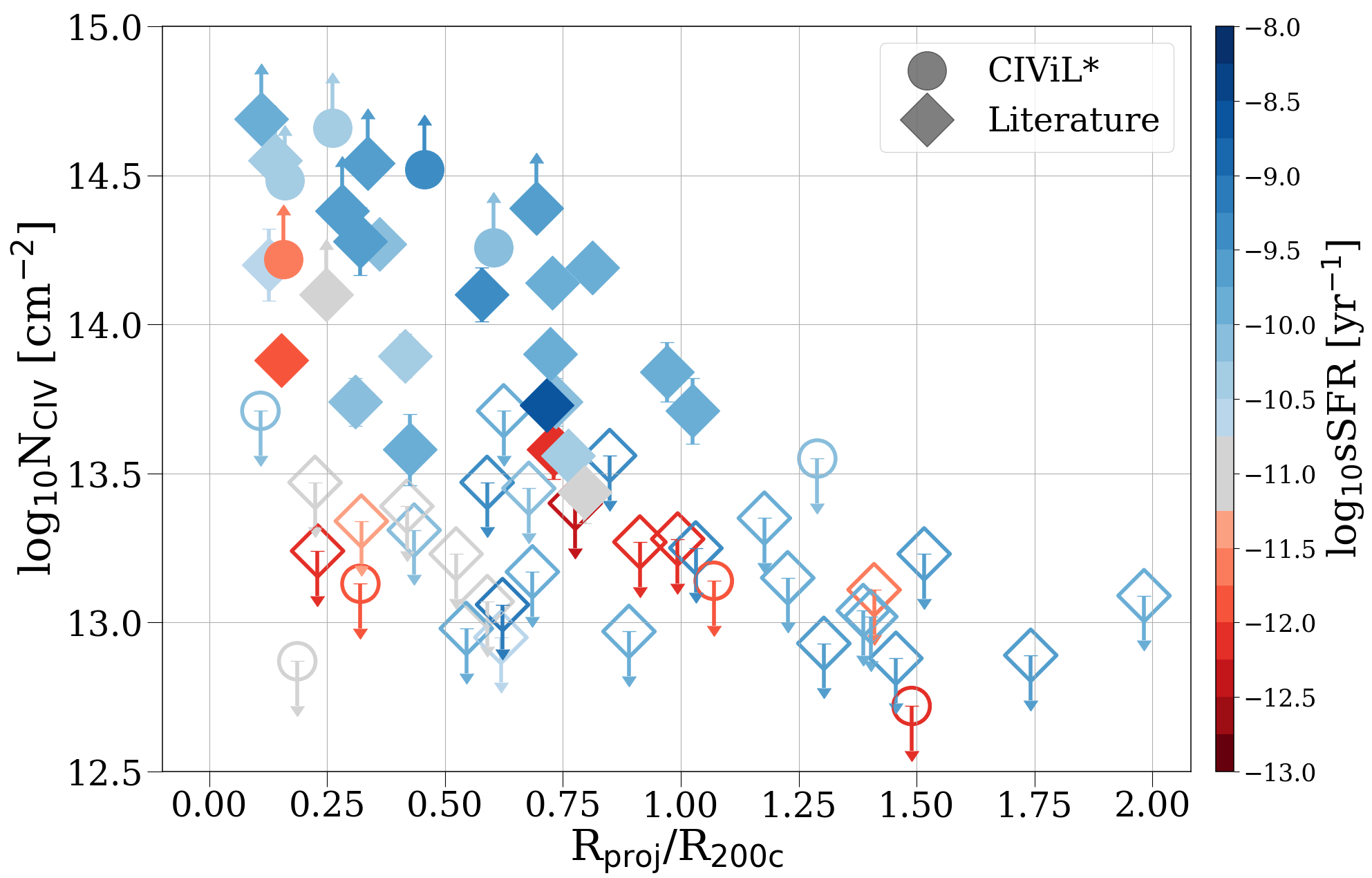}
     \qquad
     {\includegraphics[width= 0.96\columnwidth]{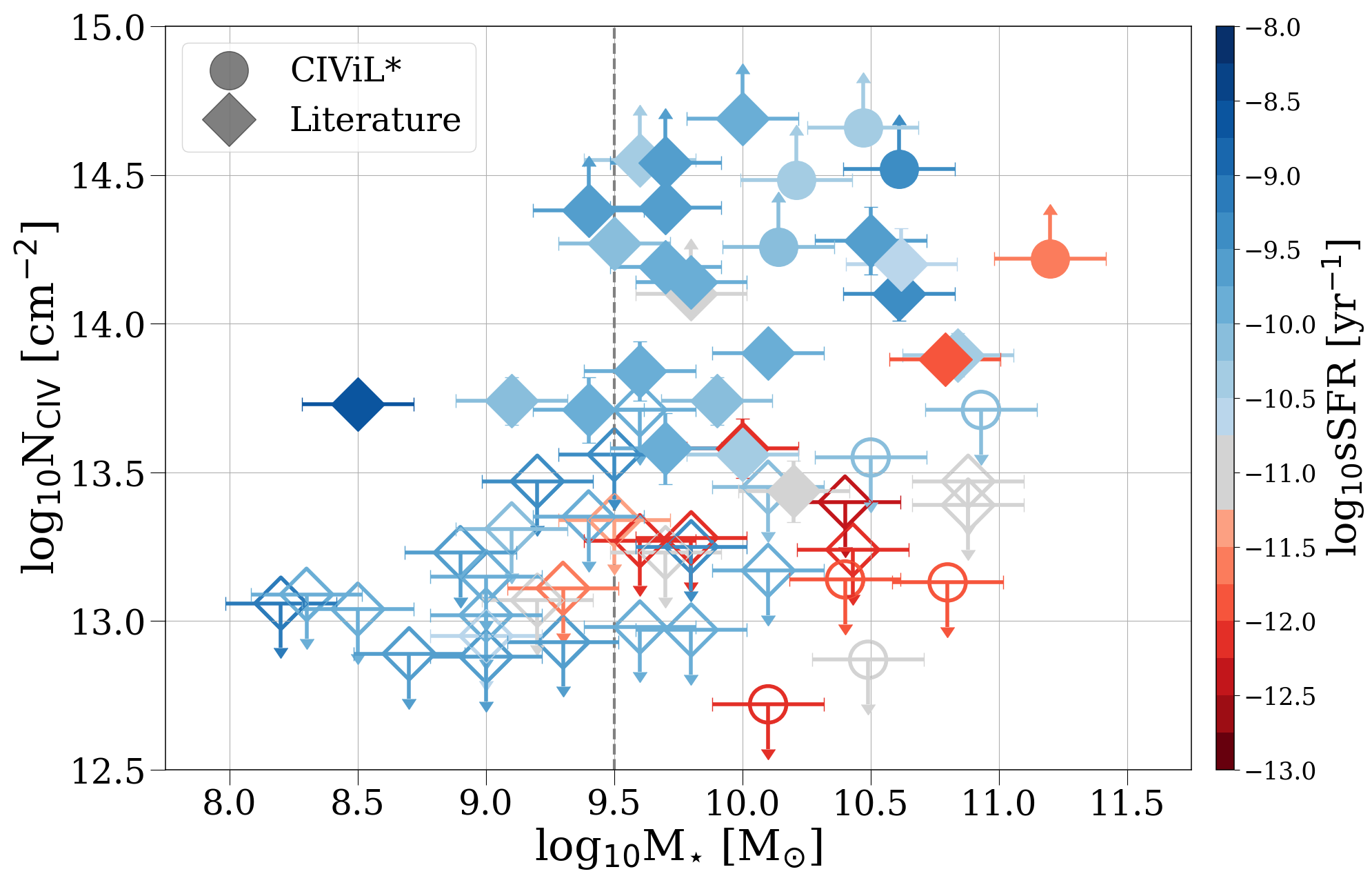}}
    \caption{{\bf Top Panel:} The C\ion{iv} detection fraction vs. normalized impact parameter. The shaded areas represent 2$\sigma$ Wilson Binomial Confidence Intervals across equal width radial bins. Upper limits exceeding the threshold (log$_{10}$N$_{\rm CIV}$/cm$^{-2}$ = 13.5) are excluded from the analysis. {\bf Middle and Bottom Panels:} C\ion{iv} column densities assembled from previous QSO absorption line surveys probing the CGM of low-z, galaxies (diamonds) with the new CIViL$^{\star}$ observations (circles) versus galaxy properties (middle:  R$_{\rm proj}$/R$_{\rm 200c}$; bottom: log$_{10}$M$_{\star}$/M$_{\odot}$). Each observation is colored by its corresponding sSFR determined from a combination of emission-line spectroscopy and broadband photometry. 
    Non-detections (upper limits) are represented with open symbols and arrows pointing down while saturated detections (lower limits) are represented as colored symbols with arrows pointing up. 
    For the rest of the analysis we only use observations with a log$_{10}$M$_{\star}$/M$_{\odot}$ $\geq$ 9.5.}
    \label{fig: nciv_rproj_mstar}
\end{figure}

\begin{figure*}[ht!]
\centering{\includegraphics[width=0.95\textwidth]{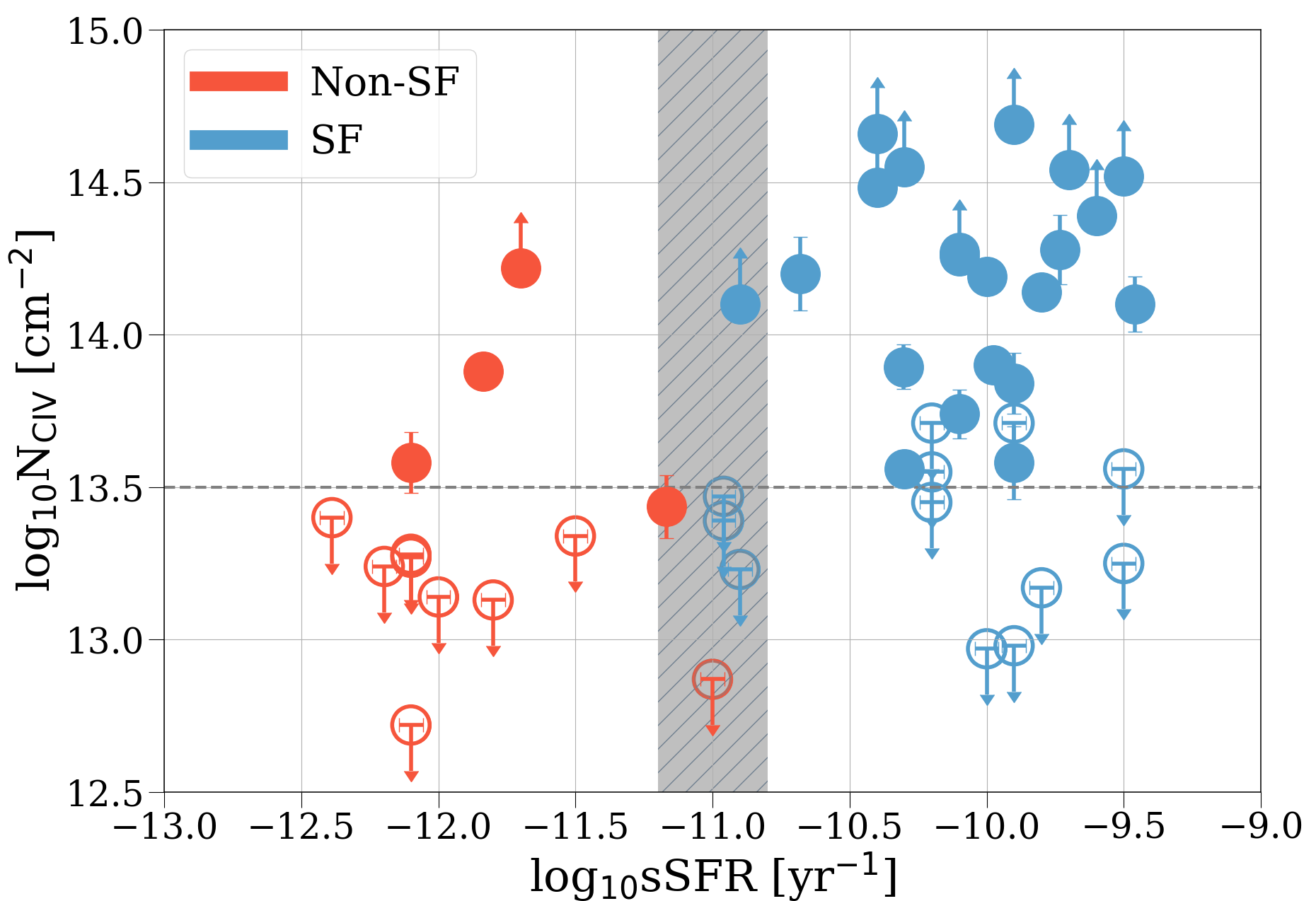}}
\caption{Measured C\ion{iv} column densities versus sSFR for CIViL$^{\star}$ and the additional literature sample, with star forming galaxies (sSFR $>$ 10$^{-11}$ yr$^{-1}$) are colored in blue while passive galaxies (sSFR $\leq$ 10$^{-11}$ yr$^{-1}$) are colored in red. For galaxies in the grey shaded area, we examine their spectra and morphology, in addition to sSFR for classification. Like the previous figure, non-detections (upper limits) are represented with open symbols and arrows pointing down while saturated detections (lower limits) are represented as colored symbols with arrows pointing up. The horizontal line signifies the detection limit of the sample. Above log$_{10}$N$_{\rm CIV}$/cm$^{-2}$ = 13.5, the star forming and passive galaxy sample have detection fractions of 72$_{-18}^{+14}$\% [21/29] and 23$_{-15}^{+27}$\% [3/13], respectively. Upper limits exceeding the threshold are excluded from the analysis. 
\label{fig: nciv_vs_sSFR_cf}}
\end{figure*}

\begin{figure*}[ht!]
\centering
   {{\includegraphics[width=0.78\textwidth]{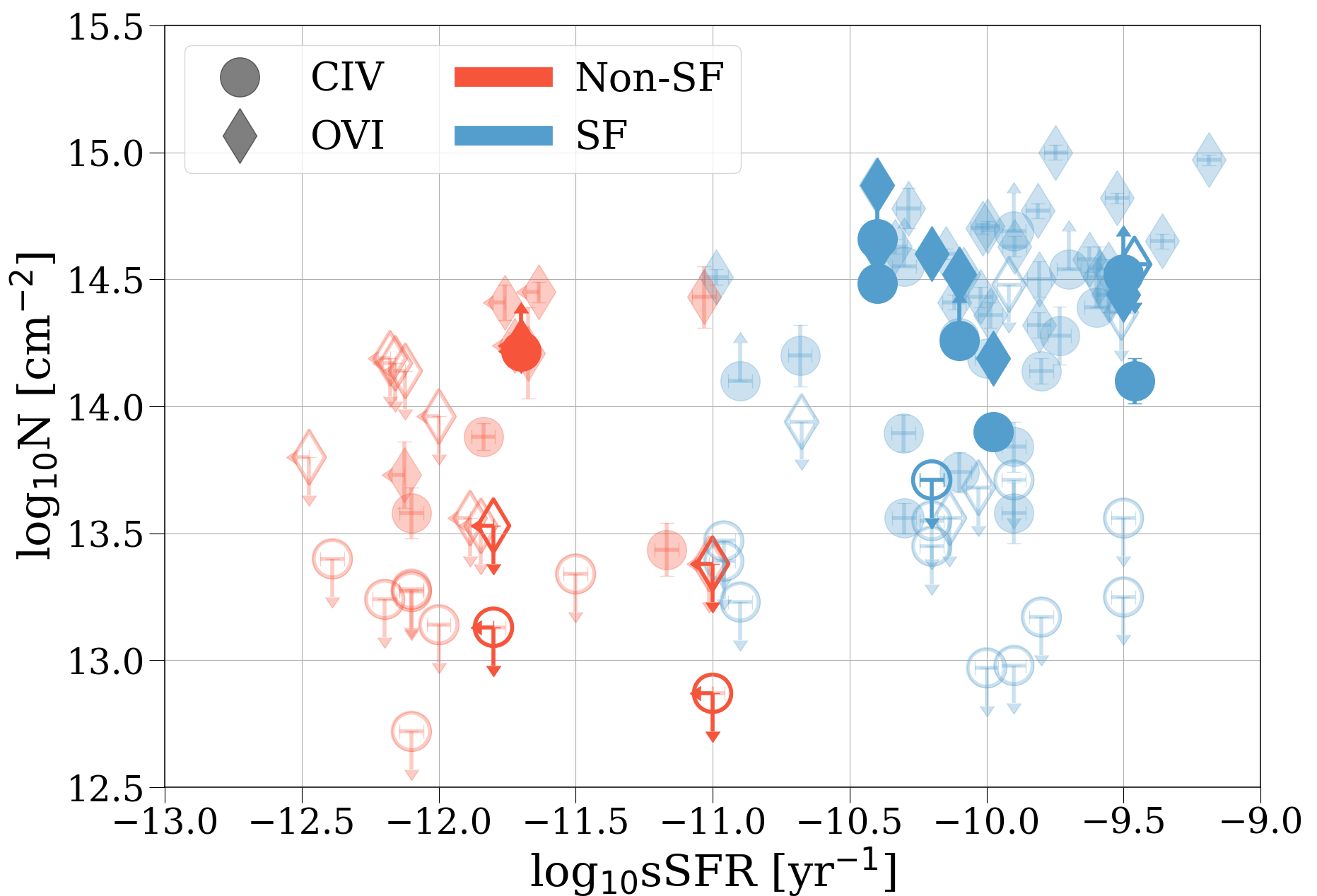}}}%
   \quad
   {{\includegraphics[width=0.78\textwidth]{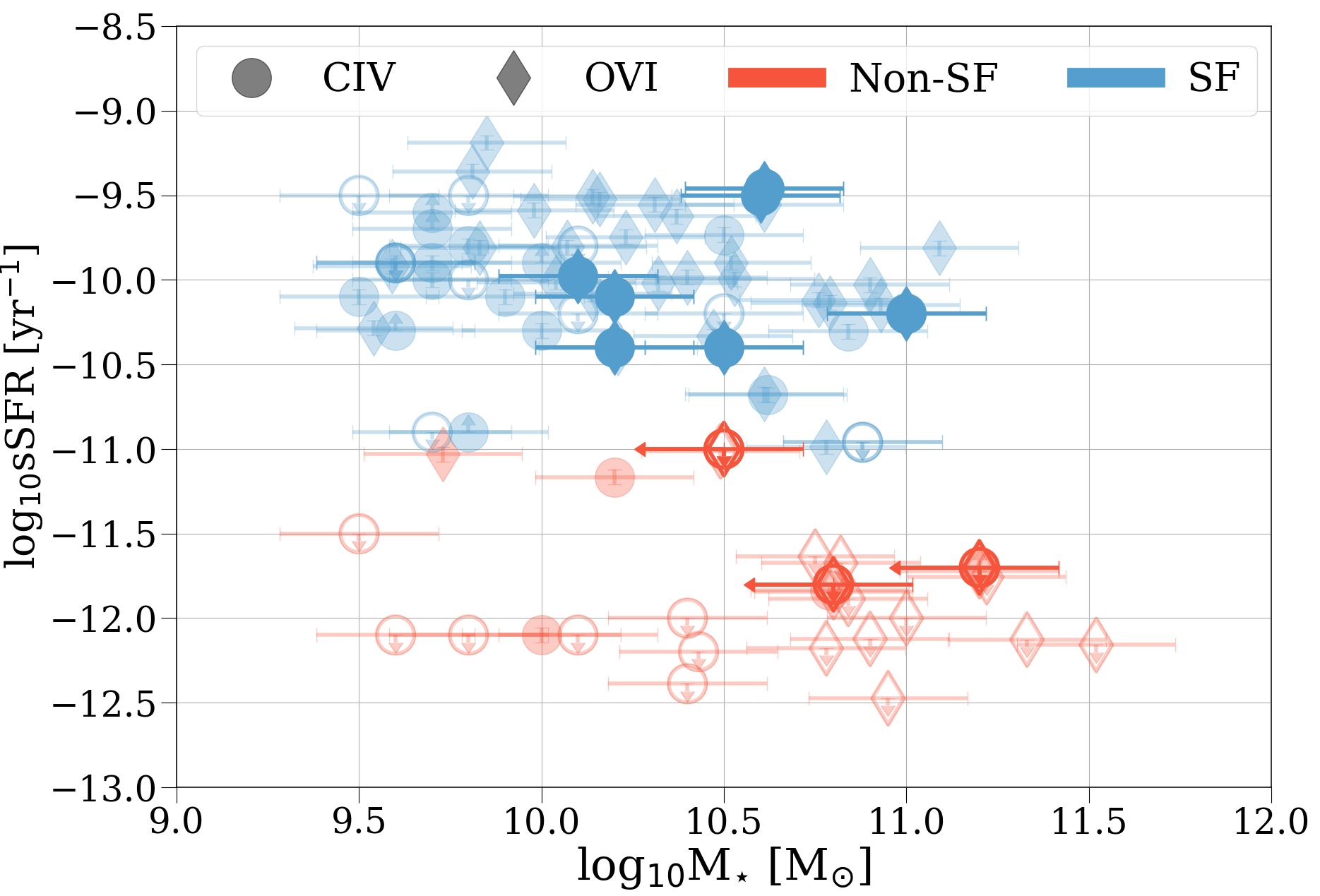}}}%
\caption{C\ion{iv} and O\ion{vi} correlations with galaxy properties. If the marker is dark/ bold, they are from galaxies where both ions were observed. These observations overlay lightly shaded O\ion{vi} column densities from \cite{Tumlinson_2011} and \cite{werk_2013} and C\ion{iv} observations from the combined CIViL$^{\star}$ and literature sample. Similar coloring and limit convention as Figure \ref{fig: nciv_vs_sSFR_cf}. {\bf Top:} log$_{10}$N$_{\rm CIV}$/cm$^{-2}$ and log$_{10}$N$_{\rm OVI}$/cm$^{-2}$ vs log$_{10}$sSFR/yr$^{-1}$. {\bf Bottom:} log$_{10}$N$_{\rm CIV}$/cm$^{-2}$ and log$_{10}$N$_{\rm OVI}$/cm$^{-2}$ vs log$_{10}$M$_{\star}$M$_{\odot}$. The basic dichotomy observed between star-forming (``blue cloud") and passive (``red sequence") galaxies seen in O\ion{vi} observations \citep{Tumlinson_2011, Johnson_2015, Zahedy_2019, Tchernyshyov_2023} is also seen in observations of C\ion{iv}.  
}
\end{figure*}

In the top panel of Figure \ref{fig: nciv_rproj_mstar} we show the C\ion{iv} detection fractions (C$_{f}$), with 2$\sigma$ Wilson Binomial confidence intervals\footnote{This method of calculating confidence intervals is better than normal approximations since it is asymmetric and can be used with small samples and skewed observations.} for 8 equal sized radial bins. Our detection fractions show that as the normalized impact parameter increases, the detection fraction of C\ion{iv} decreases; this declining profile, a trend described in several previous works \citep[e.g.][]{Bordoloi_2014}, is mirrored in the middle panel\footnote{In Appendix \ref{a: det_frac} we examine other binning methods and find these also result in a covering fraction that declines with the impact parameter.}. In the middle and bottom panel of Figure \ref{fig: nciv_rproj_mstar} we show the C\ion{iv} column densities versus impact parameter normalized by virial radius or R$_{\rm proj}$/R$_{\rm 200c}$ (middle panel) and stellar mass, log$_{10}$M$_{\star}$/M$_{\odot}$, (bottom panel) colored by specific star-formation rate (log$_{10}$sSFR/yr$^{-1}$). 

Given the variation of CGM columns with galaxy mass and redshift found in previous studies \citep{Tchernyshyov_2023}, for the rest of our analysis we restrict the sample to log$_{10}$M$_{\star}$/M$_{\odot}$ $\geq$ 9.5, since below this limit the vast majority of galaxies are star forming. Additionally, stellar masses lower than this cutoff are mostly dwarf galaxies where it has been shown than their CGM contains $\sim$10\% of metals in the cool phase (T $\approx$ 10$^{4}$ K) \citep{Zheng_2024}. This is consistent with what we see in our lower mass sample, where most of the observations of C\ion{iv} are upper limits. Introducing this cut leaves us with a sample of 46 observations of galaxies with redshifts between 0.0010 and 0.252, impact parameters of 14 to 224 kpc, and stellar masses between 10$^{9.5}$ and 10$^{11.2}$ M$_{\odot}$.

\subsection{CIV-sSFR Dichotomy}\label{sec: civ_dichotomy}

We now examine the relation between C\ion{iv} and sSFR to investigate whether a similar bimodality exists like the one using O\ion{vi} \citep{Kauffmann_2003, Schiminovich_2007, Tumlinson_2011, Tchernyshyov_2023}. 
We plot log$_{10}$N$_{\rm CIV}$/cm$^{-2}$ versus log$_{10}$sSFR/yr$^{-1}$ for the combined small sample in Figure \ref{fig: nciv_vs_sSFR_cf}. We separate the observations into two galaxy groups based on cuts made in previous works \citep{Tumlinson_2011, Tchernyshyov_2023}, divided into star-forming galaxies (sSFR $>$ 10$^{-11}$ yr$^{-1}$, blue) and passive galaxies (sSFR $\leq$ 10$^{-11}$ yr$^{-1}$, red). We paid special attention to galaxies within $\pm$ 0.2 dex of this cutoff, represented as the grey shaded area in Figure \ref{fig: nciv_vs_sSFR_cf}. Galaxies in this area are possibly transitioning between star-forming and passive and have subtleties that define their classification that a simple sSFR cut doesn't catch. For the galaxies that fall within this ``grey" area, we referred to their optical spectra and  morphology to determine their classification; for more details we refer the reader to Appendix \ref{a: grey}. 

We find detection fractions of 72$_{-18}^{+14}$\% [21/29] for the star-forming sample and 23$_{-15}^{+27}$\% [3/13] for the passive sample with 2$\sigma$ Wilson Binomial confidence intervals above log$_{10}$N$_{\rm CIV}$/cm$^{-2}$ = 13.5\footnote{See Appendix \ref{a: sf_p_det_fracs} for an exploration of detection fractions using different detections thresholds}. Using {\tt\string scipy.stats.anderson\_ksamp}\footnote{https://github.com/scipy/scipy}, we perform an Anderson-Darling test for k-samples\footnote{The k-sample Anderson-Darling test is a modification of the one-sample Anderson-Darling test. It tests the null hypothesis that k-samples are drawn from the same population without having to specify the distribution of that population.} 
to compare the C\ion{iv} column densities of the star-forming (33) and passive (13) sample. For this test we treat each observation as if it were a detection (i.e., not taking into account upper and lower limits); the Anderson-Darling test is useful when not taking into account known limits since the test is designed to examine the tails of distributions which makes it particularly sensitive to deviations in extreme values. We find that we can reject the null hypothesis that these two samples were drawn from the same distribution at a $>$ 99.5\% confidence level (p = 0.0016). We also explore statistical methods that take into account censored data (upper and lower limits) in Appendix \ref{a: alternative_stat_methods}; both of these alternative methods (interval-censored analysis: p = 0.017, two-sided k-sample test: p = 0.034) confirm our original finding that there is a bimodality between the C\ion{iv} content in star-forming and passive galaxies with $>$2$\sigma$ confidence. 

We also test the statistical significance of the correlation of N$_{\rm CIV}$ and sSFR by performing a generalized Kendall's tau test using {\tt\string scipy.stats.kendalltau}\footnote{https://github.com/scipy/scipy}. This test, which did not take into account upper or lower limits resulted in a $\tau$ = 0.29 (p = 0.004) which leads us to reject the null hypothesis that there is no correlation between the C\ion{iv} absorption and galaxy sSFR. 
\cite{Bordoloi_2014} report a similar result for galaxies with 9 $\leq$ log$_{10}$M$_{\star}$/M$_{\odot}$ $\leq$ 10 where they also rejected the null hypothesis that star-forming and passive galaxies draw from the same parent distribution of equivalent widths at a $>$99.5\% confidence level. Similar KS and Kendall's tau tests using O\ion{vi} column densities were performed by \cite{Tumlinson_2011} where they used QSO-Galaxy pairs from the COS-Halos Survey \citep{werk_2013}. They also found that N$_{\rm OVI}$ and sSFR correlate at a $>$ 99\% confidence which led them to conclude that their results show a basic dichotomy between star-forming (``blue-cloud") and passive (``red-sequence") galaxies is present in the gaseous halos of their sample. 

To compare between our result and the O\ion{vi}-sSFR dichotomy, we recreate Fig. 3 from \cite{Tumlinson_2011}. We highlight galaxies in which C\ion{iv} and O\ion{vi} were both observed in dark, bold points. These observations overlay the lightly shaded O\ion{vi} column densities from \cite{Tumlinson_2011} and \cite{werk_2013} and C\ion{iv} observations from the combined CIViL$^{\star}$ and literature sample. 
Our results independently suggest a  dichotomy between the halos of star-forming and passive galaxies as traced by C\ion{iv} but overlaid with O\ion{vi}, they statistically mirror the dichotomy discussed in \cite{Tumlinson_2011}. 

\subsection{Minimum Mass of Carbon in the CGM}\label{sec: min_mass_carbon}

We estimate the total mass of carbon in the CGM $\sim$L$^{\star}$ galaxies by following the method outlined in \cite{Bordoloi_2014}. By assuming a conservative ionization correction \citep[$f_{\rm C IV}$ = 0.3, for more details please see \S5 in][]{Bordoloi_2014} they obtained a lower limit on the carbon mass (M$_{\rm carbon}$) which can be written as: 
\begin{equation}
\begin{split}
    M_{ \rm carbon} \gtrsim 1.12 \times 10^6 \rm \: M_{\odot} \left(\frac{N_{\rm C IV,\:mean}}{10^{14} \: \rm cm^{-2}}\right) \\
    \times \left(\frac{ \rm R_{\rm proj}}{120\: \rm kpc}\right){^2} \times \left(\frac{0.3}{f_{\rm CIV}}\right). 
\end{split}
\end{equation}
Similarly to \cite{Bordoloi_2014}, we split our sample into three radial bins and then summed them to obtain the final lower limit on carbon mass. The mean column density within R$_{\rm proj}$ $<$ 40 kpc is 2.10 $\times$ 10$^{14}$ cm$^{-2}$, within 40 kpc $\leq$ R$_{\rm proj}$ $<$ 80 kpc is 1.44 $\times$ 10$^{14}$ cm$^{-2}$, and within 80kpc $\leq$ R$_{\rm proj}$ $<$ 120 kpc is 1.65 $\times$ 10$^{14}$ cm$^{-2}$. 

We find a minimum carbon mass of M$_\mathrm{carbon}$/M$_{\odot} \gtrsim$ 3.03 $\times$ $10^{6}$. We note that the column densities derived using Voigt profile fits are saturated for some of the lines of sight in the sample and are thus probably underestimating the true column density; however, this is value is about a factor of 1.6 higher than the M$_{\rm carbon}$ value presented in \cite{Bordoloi_2014} (1.9$\times 10^{6}$ M$_{\odot}$). Our minimum carbon mass is comparable to the total carbon mass in the ISM of L$^{\star}$ galaxies, \citep[e.g.][]{Peeples_2014, Bordoloi_2014}. 


\section{Discussion}\label{sec: discussion}

Results from the COS-Dwarfs survey \citep{Bordoloi_2014} suggested a correlation between C\ion{iv} absorption strength and sSFR of sub-L$^{\star}$ galaxies within half the virial radius. The COS-Holes survey \citep{Garza_2024} tentatively confirmed with $\gtrsim2\sigma$ significance that a correlation between sSFR and C\ion{iv}-bearing CGM in L$^{\star}$ galaxies is similar to that of O\ion{vi}. 
The combined CIViL$^{\star}$ observations with those from other COS-CGM surveys indicate that C\ion{iv} is more O\ion{vi}-like than ``low-ion-like". This suggests that C\ion{iv} is also tracing gas formed or maintained by star formation and/or feedback unlike other low-ionization state gas traced by singly and doubly ionized species (e.g. H\ion{i}, Si\ion{ii}, C\ion{iii}, etc) which show no correlation with galaxy star-forming properties \citep{werk_2013}. In an upcoming paper, we will perform a detailed analysis of the kinematics and ionization state of C\ion{iv}-bearing gas to provide more complete constraints on the physical conditions of the CGM of $\sim$L$^{\star}$ galaxies.   



CGM C\ion{iv} also provides a potential avenue for exploring how galaxies sustain their star formation since the CGM is a large gaseous reservoir and a source for the galaxy's star-forming fuel \citep{tumlinson_2017}. To address this, we estimate the depletion time, $\tau_{\rm dep}$, as the CGM mass divided by the mean SFR, as the timescale over which star-formation could be maintained its current rate, given an available gas supply, and assuming no inflows of fresh fuel or recycling of the gas \citep{saintonge_catinella_2022}. But how much fuel is actually available? To do this back of the envelope calculation, we use the minimum mass of carbon we estimate in \S\ref{sec: min_mass_carbon} (for a conservative ionization correction) and translate it to a total hydrogen mass using a metallicity of $Z = 1/3$ \citep{prochaska_2017_metallicity} and the solar carbon abundance. This gives a lower limit of M$_{\rm CGM}$ $\geq$ 2.83 $\times$ 10$^{9}$ \rm M$_{\odot}$ on the total gas mass in the CGM. Using the mean SFR in our sample, $1.5~{\rm M_{\odot}/yr}$, the resulting lower limit on the depletion time is $\tau _{\rm dep} \geq$ 1.93 Gyr.


Comparing our depletion time to depletion times presented in Fig. 6a of \cite{peroux_howk_2020}, we find that our $\tau_{\rm dep}$ is fairly consistent with their molecular gas depletion timescale at $z$=0. Our depletion timescale is comparable or shorter than the dynamical timescale (taken to be 10\% of the Hubble time) which suggests that the CGM and molecular gas available to galaxies, assuming no inflows of fresh fuel or recycling of the gas, slowly becomes insufficient to fuel star formation on its own \citep{peroux_howk_2020}. Therefore, galaxies are most likely undergoing some resupply process either through the conversion of ionized gas from the CGM or IGM or through accretion from the IGM onto the disks of galaxies. Thus, gas in the CGM is only one piece of the large reservoir that galaxies use as fuel for future star formation. 





\section{Summary and Conclusion}\label{sec: summary_and_conclusion}

In this work, we presented observations from the CIV in L$^{\star}$ galaxies (CIViL$^{\star}$) Survey. The main results of this study are as follows. 

\begin{enumerate}
    \item The CIViL$^{\star}$ Survey amplifies the diagnostic power of the current COS-CGM samples through the acquisition of 11 new C\ion{iv} observations for L $^{\star}$ galaxies. 
    
    \item We combine our observations from CIViL$^{\star}$ with C\ion{iv} observations \citep{werk_2013, Bordoloi_2014, Garza_2024} for a final sample of 45 lines of sight. We separate the observations by their sSFR where we classify them as either star-forming (sSFR $>$ 10$^{-11}$ yr$^{-1}$) or passive (sSFR $\leq$ 10$^{-11}$ yr$^{-1}$). 
    
    \item We find a detection fraction of 72$_{-18}^{+14}$\% [21/29] for the star-forming sample and 23$_{-15}^{+27}$\% [3/13] for the passive sample. Using an Anderson-Darling test to compare C\ion{iv} absorption in star-forming and passive galaxies, we find a dichotomy at a $>$99.5\% confidence level. 
    
    \item Our discovery of a dichotomy in L$^{\star}$ galaxies is similar to the one found using O\ion{vi} bearing gas \citep{Tumlinson_2011, Tchernyshyov_2023}.   

    
\end{enumerate}

The results from this paper are the tip of the iceberg for what observations from the CIViL$^{\star}$ Survey will reveal. This survey supplements the enormous investment of 483 orbits in previous COS-CGM surveys by placing one of the most consistent ion tracers of diffuse gas in the context of the baryon cycle over 10+ Gyrs of cosmic evolution. Not only does it close a gap in C\ion{iv} coverage for low-$z$, L$^{\star}$ galaxies it provides the opportunity to constrain how the baryon cycle differs among dwarf, star-forming, passive, and AGN-bearing galaxies. Future work carried out using the CIViL$^{\star}$ survey data will examine the kinematics and ionization mechanisms of the C\ion{iv}-traced gas phase of the CGM and the differences between the CGM of AGN hosts and star-forming galaxies.

\begin{acknowledgments}
Based on observations with the NASA/ESA Hubble Space Telescope through program number HST-GO-17076, obtained at the Space Telescope Science Institute, which is operated by the Association of Universities for Research in Astronomy, Incorporated, under NASA contract NAS5-26555. Additional data comes from HST-GO programs: COS-Halos PID$\#$11598, COS-Dwarfs PID$\#$12248, COS-Holes PID$\#$16650. This research has made use of the NASA/IPAC Extragalactic Database (NED), which is operated by the Jet Propulsion Laboratory, California Institute of Technology, under contract with the National Aeronautics and Space Administration. 

We thank the anonymous referee for their insightful comments that improved the quality of the paper. SLG recognizes the unceded traditional lands of the Duwamish and Puget Sound Salish Tribes, on which she is grateful to love and work. They also thank Boeun Choi and Kirill Tchernyshyov for enlightening conversations and advice that helped improve this work. YF acknowledges support by NASA award 19-ATP19-0023 and NSF award AST-2007012.
\end{acknowledgments}

%

\vspace{5mm}
\facilities{HST(COS)}

Data Availability: {\it HST}/COS spectra can be found on in MAST: \dataset[doi:10.17909/dgjk-f804]{http://dx.doi.org/10.17909/dgjk-f804}.


\software{astropy \citep{astropy_2013,astropy_2018},  
          Cloudy \citep{ferland_2013_cloudy}, 
          {\tt interval} \citep{fay2010exact},
          {\tt linetools} \citep{prochaska_linetools_2017}, {\tt matplotlib} \citep{Hunter:2007}, 
          {\tt numpy} \citep{harris2020array},
          {\tt pandas} \citep{pandas_2024},
          {\tt PyIGM} \citep{prochaska_pyigm_2017},
          {\tt scipy} \citep{2020SciPy-NMeth},
          {\tt veeper} \citep{burchett_veeper_2024}.} 




\appendix

\section{Full Sample Detection Fractions}\label{a: det_frac}

\setcounter{table}{0}
\renewcommand{\thetable}{\Alph{section}.\arabic{table}}
\renewcommand*{\theHtable}{\thetable}

In addition to calculating covering fractions for equal radial bins, we created radial bins that had an equal amount of galaxies per bin. We list detection fractions from both of these scenarios in Table \ref{tab: det_fracs}. Again, our detection fractions reflect the declining radial profile seen in the middle panel of Figure \ref{fig: nciv_rproj_mstar}. 

\begin{table*}[]
\centering
\caption{Full Sample Detection Fractions}
\begin{tabular}{cccc}
\hline
bin width           & hit rate           & C$_{f}$          & 2$\sigma$ CI           \\
(1)                 & (2)                & (3)              & (4)                    \\ \hline
\multicolumn{4}{c}{4 Equal Radial Bins (R$_{\rm proj}$/R$_{\rm 200c}$)}                \\ \hline
0.0-0.5             & 16/23              & 0.70            & (0.49, 0.84)         \\
0.5-1.0             & 11/24              & 0.46            & (0.28, 0.65)         \\
1.0-1.5             & 1/11               & 0.09            & (0.02, 0.38)         \\
1.5-2.0             & 0/3                & 0.00            & (0, 0)                 \\ \hline
\multicolumn{4}{c}{8 Equal Radial Bins (R$_{\rm proj}$/R$_{\rm 200c}$)}                \\ \hline
0.0-0.25             & 7/10              & 0.70            & (0.40, 0.89)         \\
0.25-0.5             & 9/13              & 0.69            & (0.42, 0.87)         \\
0.5-0.75             & 8/16              & 0.50            & (0.28, 0.72)         \\
0.75-1.0             & 3/8              & 0.38            & (0.14, 0.69)         \\
1.0-1.25             & 1/5               & 0.20            & (0.04, 0.62)         \\
1.25-1.5             & 0/7                & 0.00            & (0, 0)
\\
1.5-1.75             & 0/2                & 0.00            & (0, 0)
\\
1.75-1.0             & 0/1                & 0.00            & (0, 0)
\\ \hline
\multicolumn{4}{c}{Equal Number of Galaxies Per Bin (R$_{\rm proj}$/R$_{\rm 200c}$)} \\ \hline
0.1-0.3             & 9/12               & 0.75            & (0.47, 0.91)         \\
0.3-0.5             & 7/13               & 0.54            & (0.29, 0.77          \\
0.6-0.7             & 6/12               & 0.50            & (0.25, 0.75)         \\
0.7-1.0             & 6/12               & 0.50            & (0.25, 0.75)         \\
1.0-2.0             & 0/12               & 0.00            & (0, 0)                 \\ \hline
\end{tabular}
\label{tab: det_fracs}
\tablecomments{Comments on columns: (1) width of the radial bins; (2) hit rate - the amount of detections above the threshold of log$_{10}$N$_{CIV}$/cm$^{-2}$ = 13.5 (3) detection fractions, (4) 2$\sigma$ Wilson Binomial Confidence Intervals.}
\end{table*}

\section{Galaxies in the ``Grey" Area}\label{a: grey}

There are five galaxies in our sample that have sSFRs that fall within a ``grey" area of $\pm$0.2 dex around our cut off of log$_{10}$sSFR/yr$^{-1}$ = -11.0, or the area where galaxies are transition between star-forming and passive. To classify these galaxies as either star-forming or passive, we took extra steps in addition to looking at their sSFR. Our explanation our choice to denote a galaxy as either star-forming or passive is detailed below: 

\begin{itemize}
    \item SDSS J110404.25+314015.1 \citep[COS-Dwarfs, 211\_65,][]{Bordoloi_2014}: We looked at the spectrum available through the Sloan Digital Sky Survey (SDSS) and observed strong emission lines indicative of a late type galaxy. We group this galaxy in the star-forming bin for our analysis. 

    \item SDSS J082022.99+233447.4 \citep[COS-Halos and COS-Dwarfs, 260\_17,][]{werk_2013,Bordoloi_2014}: We looked at the spectrum available through SDSS and observed both absorption and emission features. We investigated the galaxy's emission line ratios and compared them to an SDSS BPT diagram and saw what it lies within the star-forming region. In addition, this galaxy is a dusty, edge on disk galaxy with an log$_{10}$sSFR/yr$^{-1}$ = -10.9. Thus we place this galaxy in the star-forming group for our analysis. 

    \item NGC 4258 \citep[COS-Holes,][]{Garza_2024}: We looked at the spectrum available on NED\footnote{The NASA/IPAC Extragalactic Database (NED) is funded by the National Aeronautics and Space Administration and operated by the California Institute of Technology.} and observed strong emission features indicative of a late type galaxy. We also investigated the morphology of NGC 4258 and found it to be a weakly barred spiral galaxy. Due to these findings we place this galaxy in the star-forming group for our analysis. 

    \item NGC 3489 \citep[COS-Holes,][]{Garza_2024}: We looked at the spectrum available on NED and observed both emission and absorption features. Looking at the morphology we find that the galaxy is an intermediate spiral. Taking into consideration its sSFR (log$_{10}$sSFR/yr$^{-1}$ = -11.167) we group this in the passive galaxy group. 

    \item SDSS J134252.23-005343.2 \citep[COS-Halos, 77\_10,][]{werk_2013}: We looked at the spectrum available in \cite{werk_2012} and determined that it is very clearly an early type galaxy from the spectrum with distinct absorption lines. We place this galaxy in the passive group for our analysis.  
\end{itemize}

\section{Star-Forming \& Passive Detection Fractions}\label{a: sf_p_det_fracs}

\begin{table*}[]
\centering
\caption{SF \& Non-SF Detection Fractions}
\begin{tabular}{cccc}
\hline
Galaxy Group           & hit rate         & C$_{f}$         & 2$\sigma$ CI         \\
(1)                    & (2)              & (3)             & (4)                  \\ \hline
\multicolumn{4}{c}{Detection Threshold: log$_{10}$N$_{\rm CIV}$/cm$^{-2}$ = 13.5}  \\ \hline
SF                     & 21/29            & 72\%              & (54, 85)             \\
Non-SF                 & 3/13             & 23\%              & (8, 50)              \\ \hline
\multicolumn{4}{c}{Detection Threshold: log$_{10}$N$_{\rm CIV}$/cm$^{-2}$ = 13.75} \\ \hline
SF                     & 18/33            & 54\%              & (38, 70)             \\
Non-SF                 & 2/13             & 15\%              & (4, 42)              \\ \hline
\multicolumn{4}{c}{Detection Threshold: log$_{10}$N$_{\rm CIV}$/cm$^{-2}$ = 14.0}  \\ \hline
SF                     & 15/33            & 45\%              & (30, 62)             \\
Non-SF                 & 1/13             & 8\%               & (1, 33)              \\ \hline
\end{tabular}
\label{tab: sf_p_det_fracs}
\tablecomments{Comments on columns: (1) Galaxy Group - either star-forming (SF) or passive (Non-SF); (2) hit rate - the amount of detections above the indicated detection threshold (3) detection fractions, (4) 2$\sigma$ Wilson Binomial Confidence Intervals.}
\end{table*}

Here we compare the detection fractions calculated using different detection thresholds as shown in Table \ref{tab: sf_p_det_fracs}. Upper limits exceeding the threshold are excluded from the analysis, counting as neither detections nor non-detections. The probability distribution for the detection fraction is characterized by a Beta distribution:
\begin{equation}
    p(f|k,N) = Beta(\alpha = k + c, \beta = N - K +c)
\end{equation}
The likelihood of observing $k$ detections in $N$ trials given a detection probability $f$ is given by the binomial distribution. The posterior probability distribution for $f$ given an observed number of detections and assuming a Jeffreys’ prior on $f$ is a beta distribution with parameters $\alpha=k+1/2$ and $\beta=N-k+1/2$. This beta distribution arises from the normalization of a binomial probability mass function $P(k|N,f) \propto f^k(1-f)^{N-k}$ with respect to the parameter $f$ rather than the count $k$. The statistical comparison of detection fractions is performed through Monte Carlo sampling from the respective beta distributions. For each pair of covering fractions, we examine ratio distributions $f_{\rm SF}/f_{\rm P}$.  The results of these distributional comparisons across detection fraction thresholds are summarized as follows:
\begin{itemize}
    \item \textbf{Detection Threshold of log$_{10}$N$_{\rm CIV}$/cm$^{-2}$ = 13.5:} The analysis indicates that the star-forming detection fraction exceeds the passive detection fraction by a factor of $2$--$5$, measured within the $68\%$ credible interval. The probability of the passive detection fraction surpassing the star-forming detection fraction is $p = 1.2 \times 10^{-3}$, corresponding to a $3\sigma$ deviation under the assumption of normality.

    \item \textbf{Detection Threshold of log$_{10}$N$_{\rm CIV}$/cm$^{-2}$ = 13.75:} The analysis indicates that the star-forming detection fraction exceeds the passive detection fraction by a factor of $2$ to $7$, measured within the $68\%$ credible interval. The probability of the passive detection fraction surpassing the star-forming detection fraction is $p = 6.2 \times 10^{-3}$, corresponding to a $2.5\sigma$ deviation under the assumption of normality. 

    \item \textbf{Detection Threshold of log$_{10}$N$_{\rm CIV}$/cm$^{-2}$ = 14.0:} The analysis indicates that the star-forming detection fraction exceeds the passive detection fraction by a factor of $2$ to $14$, measured within the $68\%$ credible interval. The probability of the passive detection fraction surpassing the star-forming detection fraction is $p = 4.8 \times 10^{-3}$, corresponding to a $2.6\sigma$ deviation under the assumption of normality. 
\end{itemize}

\section{Statistical Methods for Censored Data}\label{a: alternative_stat_methods}

For observations characterized by detection limits, where measurements are confined by lower and upper bounds, interval-censored statistical methods provide a practical, yet valid, analytical framework. For upper limits, we construct intervals from a common lower bound to the detection threshold, while for lower limits, we construct intervals from the saturation limit to a common upper bound. The implementation of interval-censored survival analysis thus accommodates both left-censored (upper limits) and right-censored (lower limits) data by transforming point constraints into comparable intervals. This approach maintains statistical validity in rank-based tests, as the specific values chosen for the common bounds do not affect the relative ordering of the observations, provided these bounds are consistently applied across all measurements and are outside the range of the measurements.

Interval-censored survival analysis was performed using the \texttt{interval} package in R to evaluate differences between the survival distributions of the passive and star-forming samples. The log-rank test with the \texttt{logrank1} score function was implemented to test the hypothesis that the survival distribution of the passive sample is stochastically less than that of the star-forming sample. This test yielded statistical significance at $p = 0.017$. A subsequent two-sided k-sample test indicated that the survival distributions differ significantly ($p = 0.034$).
These interval-censored survival analyses, which explicitly account for the censoring structure in the observations, provide statistical evidence for bimodality in the C\ion{iv} content between star-forming and passive galaxies at $>2\sigma$ significance.

\bibliography{references}{}

\begin{thebibliography}{}
\expandafter\ifx\csname natexlab\endcsname\relax\def\natexlab#1{#1}\fi
\providecommand{\url}[1]{\href{#1}{#1}}
\providecommand{\dodoi}[1]{doi:~\href{http://doi.org/#1}{\nolinkurl{#1}}}
\providecommand{\doeprint}[1]{\href{http://ascl.net/#1}{\nolinkurl{http://ascl.net/#1}}}
\providecommand{\doarXiv}[1]{\href{https://arxiv.org/abs/#1}{\nolinkurl{https://arxiv.org/abs/#1}}}

\bibitem[{{Anand} {et~al.}(2021){Anand}, {Nelson}, \& {Kauffmann}}]{Anand_2021}
{Anand}, A., {Nelson}, D., \& {Kauffmann}, G. 2021, \mnras, 504, 65, \dodoi{10.1093/mnras/stab871}

\bibitem[{{Astropy Collaboration} {et~al.}(2013){Astropy Collaboration}, {Robitaille}, {Tollerud}, {Greenfield}, {Droettboom}, {Bray}, {Aldcroft}, {Davis}, {Ginsburg}, {Price-Whelan}, {Kerzendorf}, {Conley}, {Crighton}, {Barbary}, {Muna}, {Ferguson}, {Grollier}, {Parikh}, {Nair}, {Unther}, {Deil}, {Woillez}, {Conseil}, {Kramer}, {Turner}, {Singer}, {Fox}, {Weaver}, {Zabalza}, {Edwards}, {Azalee Bostroem}, {Burke}, {Casey}, {Crawford}, {Dencheva}, {Ely}, {Jenness}, {Labrie}, {Lim}, {Pierfederici}, {Pontzen}, {Ptak}, {Refsdal}, {Servillat}, \& {Streicher}}]{astropy_2013}
{Astropy Collaboration}, {Robitaille}, T.~P., {Tollerud}, E.~J., {et~al.} 2013, \aap, 558, A33, \dodoi{10.1051/0004-6361/201322068}

\bibitem[{{Astropy Collaboration} {et~al.}(2018){Astropy Collaboration}, {Price-Whelan}, {Sip{\H{o}}cz}, {G{\"u}nther}, {Lim}, {Crawford}, {Conseil}, {Shupe}, {Craig}, {Dencheva}, {Ginsburg}, {VanderPlas}, {Bradley}, {P{\'e}rez-Su{\'a}rez}, {de Val-Borro}, {Aldcroft}, {Cruz}, {Robitaille}, {Tollerud}, {Ardelean}, {Babej}, {Bach}, {Bachetti}, {Bakanov}, {Bamford}, {Barentsen}, {Barmby}, {Baumbach}, {Berry}, {Biscani}, {Boquien}, {Bostroem}, {Bouma}, {Brammer}, {Bray}, {Breytenbach}, {Buddelmeijer}, {Burke}, {Calderone}, {Cano Rodr{\'\i}guez}, {Cara}, {Cardoso}, {Cheedella}, {Copin}, {Corrales}, {Crichton}, {D'Avella}, {Deil}, {Depagne}, {Dietrich}, {Donath}, {Droettboom}, {Earl}, {Erben}, {Fabbro}, {Ferreira}, {Finethy}, {Fox}, {Garrison}, {Gibbons}, {Goldstein}, {Gommers}, {Greco}, {Greenfield}, {Groener}, {Grollier}, {Hagen}, {Hirst}, {Homeier}, {Horton}, {Hosseinzadeh}, {Hu}, {Hunkeler}, {Ivezi{\'c}}, {Jain}, {Jenness}, {Kanarek}, {Kendrew}, {Kern}, {Kerzendorf}, {Khvalko}, {King}, {Kirkby}, {Kulkarni},
  {Kumar}, {Lee}, {Lenz}, {Littlefair}, {Ma}, {Macleod}, {Mastropietro}, {McCully}, {Montagnac}, {Morris}, {Mueller}, {Mumford}, {Muna}, {Murphy}, {Nelson}, {Nguyen}, {Ninan}, {N{\"o}the}, {Ogaz}, {Oh}, {Parejko}, {Parley}, {Pascual}, {Patil}, {Patil}, {Plunkett}, {Prochaska}, {Rastogi}, {Reddy Janga}, {Sabater}, {Sakurikar}, {Seifert}, {Sherbert}, {Sherwood-Taylor}, {Shih}, {Sick}, {Silbiger}, {Singanamalla}, {Singer}, {Sladen}, {Sooley}, {Sornarajah}, {Streicher}, {Teuben}, {Thomas}, {Tremblay}, {Turner}, {Terr{\'o}n}, {van Kerkwijk}, {de la Vega}, {Watkins}, {Weaver}, {Whitmore}, {Woillez}, {Zabalza}, \& {Astropy Contributors}}]{astropy_2018}
{Astropy Collaboration}, {Price-Whelan}, A.~M., {Sip{\H{o}}cz}, B.~M., {et~al.} 2018, \aj, 156, 123, \dodoi{10.3847/1538-3881/aabc4f}

\bibitem[{{Bahcall} {et~al.}(1991){Bahcall}, {Jannuzi}, {Schneider}, {Hartig}, {Bohlin}, \& {Junkkarinen}}]{Bahcall_1991}
{Bahcall}, J.~N., {Jannuzi}, B.~T., {Schneider}, D.~P., {et~al.} 1991, \apjl, 377, L5, \dodoi{10.1086/186103}

\bibitem[{{Behroozi} {et~al.}(2019){Behroozi}, {Wechsler}, {Hearin}, \& {Conroy}}]{Behrooz_2019}
{Behroozi}, P., {Wechsler}, R.~H., {Hearin}, A.~P., \& {Conroy}, C. 2019, \mnras, 488, 3143, \dodoi{10.1093/mnras/stz1182}

\bibitem[{{Berg} {et~al.}(2018){Berg}, {Ellison}, {Tumlinson}, {Oppenheimer}, {Horton}, {Bordoloi}, \& {Schaye}}]{berg_2018}
{Berg}, T. A.~M., {Ellison}, S.~L., {Tumlinson}, J., {et~al.} 2018, \mnras, 478, 3890, \dodoi{10.1093/mnras/sty962}

\bibitem[{{Bergeron}(1986)}]{Bergeron_1986}
{Bergeron}, J. 1986, \aap, 155, L8

\bibitem[{{Bordoloi} {et~al.}(2018){Bordoloi}, {Prochaska}, {Tumlinson}, {Werk}, {Tripp}, \& {Burchett}}]{Bordoloi_2018}
{Bordoloi}, R., {Prochaska}, J.~X., {Tumlinson}, J., {et~al.} 2018, \apj, 864, 132, \dodoi{10.3847/1538-4357/aad8ac}

\bibitem[{{Bordoloi} {et~al.}(2011){Bordoloi}, {Lilly}, {Knobel}, {Bolzonella}, {Kampczyk}, {Carollo}, {Iovino}, {Zucca}, {Contini}, {Kneib}, {Le Fevre}, {Mainieri}, {Renzini}, {Scodeggio}, {Zamorani}, {Balestra}, {Bardelli}, {Bongiorno}, {Caputi}, {Cucciati}, {de la Torre}, {de Ravel}, {Garilli}, {Kova{\v{c}}}, {Lamareille}, {Le Borgne}, {Le Brun}, {Maier}, {Mignoli}, {Pello}, {Peng}, {Perez Montero}, {Presotto}, {Scarlata}, {Silverman}, {Tanaka}, {Tasca}, {Tresse}, {Vergani}, {Barnes}, {Cappi}, {Cimatti}, {Coppa}, {Diener}, {Franzetti}, {Koekemoer}, {L{\'o}pez-Sanjuan}, {McCracken}, {Moresco}, {Nair}, {Oesch}, {Pozzetti}, \& {Welikala}}]{Bordoloi_2011}
{Bordoloi}, R., {Lilly}, S.~J., {Knobel}, C., {et~al.} 2011, \apj, 743, 10, \dodoi{10.1088/0004-637X/743/1/10}

\bibitem[{{Bordoloi} {et~al.}(2014){Bordoloi}, {Tumlinson}, {Werk}, {Oppenheimer}, {Peeples}, {Prochaska}, {Tripp}, {Katz}, {Dav{\'e}}, {Fox}, {Thom}, {Ford}, {Weinberg}, {Burchett}, \& {Kollmeier}}]{Bordoloi_2014}
{Bordoloi}, R., {Tumlinson}, J., {Werk}, J.~K., {et~al.} 2014, \apj, 796, 136, \dodoi{10.1088/0004-637X/796/2/136}

\bibitem[{{Borthakur} {et~al.}(2015){Borthakur}, {Heckman}, {Tumlinson}, {Bordoloi}, {Thom}, {Catinella}, {Schiminovich}, {Dav{\'e}}, {Kauffmann}, {Moran}, \& {Saintonge}}]{Borthakur_2015}
{Borthakur}, S., {Heckman}, T., {Tumlinson}, J., {et~al.} 2015, \apj, 813, 46, \dodoi{10.1088/0004-637X/813/1/46}

\bibitem[{Burchett(2024)}]{burchett_veeper_2024}
Burchett, J. 2024, {the Veeper}, v1.0,  Zenodo, \dodoi{10.5281/zenodo.10993984}

\bibitem[{{Burchett} {et~al.}(2016){Burchett}, {Tripp}, {Bordoloi}, {Werk}, {Prochaska}, {Tumlinson}, {Willmer}, {O'Meara}, \& {Katz}}]{Burchett_2016}
{Burchett}, J.~N., {Tripp}, T.~M., {Bordoloi}, R., {et~al.} 2016, \apj, 832, 124, \dodoi{10.3847/0004-637X/832/2/124}

\bibitem[{{Chen} {et~al.}(2001){Chen}, {Lanzetta}, \& {Webb}}]{Chen_2001}
{Chen}, H.-W., {Lanzetta}, K.~M., \& {Webb}, J.~K. 2001, \apj, 556, 158, \dodoi{10.1086/321537}

\bibitem[{{Danforth} {et~al.}(2016){Danforth}, {Keeney}, {Tilton}, {Shull}, {Stocke}, {Stevans}, {Pieri}, {Savage}, {France}, {Syphers}, {Smith}, {Green}, {Froning}, {Penton}, \& {Osterman}}]{danforth_2016}
{Danforth}, C.~W., {Keeney}, B.~A., {Tilton}, E.~M., {et~al.} 2016, \apj, 817, 111, \dodoi{10.3847/0004-637X/817/2/111}

\bibitem[{Fay \& Shaw(2010)}]{fay2010exact}
Fay, M.~P., \& Shaw, P.~A. 2010, Journal of Statistical Software, 36, 1, \dodoi{10.18637/jss.v036.i02}

\bibitem[{{Ferland} {et~al.}(2013){Ferland}, {Porter}, {van Hoof}, {Williams}, {Abel}, {Lykins}, {Shaw}, {Henney}, \& {Stancil}}]{ferland_2013_cloudy}
{Ferland}, G.~J., {Porter}, R.~L., {van Hoof}, P.~A.~M., {et~al.} 2013, \rmxaa, 49, 137.
\newblock \doarXiv{1302.4485}

\bibitem[{{Froning} \& {Green}(2009)}]{Froning_and_Green_2009}
{Froning}, C.~S., \& {Green}, J.~C. 2009, \apss, 320, 181, \dodoi{10.1007/s10509-008-9758-y}

\bibitem[{{Garza} {et~al.}(2024){Garza}, {Werk}, {Oppenheimer}, {Tchernyshyov}, {Sanchez}, {Faerman}, {Rubin}, {Bentz}, {Davies}, {Burchett}, {Crain}, \& {Prochaska}}]{Garza_2024}
{Garza}, S.~L., {Werk}, J.~K., {Oppenheimer}, B.~D., {et~al.} 2024, \apj, 970, 115, \dodoi{10.3847/1538-4357/ad4ecc}

\bibitem[{{Green} {et~al.}(2012){Green}, {Froning}, {Osterman}, {Ebbets}, {Heap}, {Leitherer}, {Linsky}, {Savage}, {Sembach}, {Shull}, {Siegmund}, {Snow}, {Spencer}, {Stern}, {Stocke}, {Welsh}, {B{\'e}land}, {Burgh}, {Danforth}, {France}, {Keeney}, {McPhate}, {Penton}, {Andrews}, {Brownsberger}, {Morse}, \& {Wilkinson}}]{Green_2012}
{Green}, J.~C., {Froning}, C.~S., {Osterman}, S., {et~al.} 2012, \apj, 744, 60, \dodoi{10.1088/0004-637X/744/1/6010.1086/141956}

\bibitem[{Harris {et~al.}(2020)Harris, Millman, van~der Walt, Gommers, Virtanen, Cournapeau, Wieser, Taylor, Berg, Smith, Kern, Picus, Hoyer, van Kerkwijk, Brett, Haldane, del R{\'{i}}o, Wiebe, Peterson, G{\'{e}}rard-Marchant, Sheppard, Reddy, Weckesser, Abbasi, Gohlke, \& Oliphant}]{harris2020array}
Harris, C.~R., Millman, K.~J., van~der Walt, S.~J., {et~al.} 2020, Nature, 585, 357, \dodoi{10.1038/s41586-020-2649-2}

\bibitem[{{Heckman} {et~al.}(2017){Heckman}, {Borthakur}, {Wild}, {Schiminovich}, \& {Bordoloi}}]{Heckman_2017}
{Heckman}, T., {Borthakur}, S., {Wild}, V., {Schiminovich}, D., \& {Bordoloi}, R. 2017, \apj, 846, 151, \dodoi{10.3847/1538-4357/aa80dc}

\bibitem[{{Hu} \& {Kravtsov}(2003)}]{Hu_2003}
{Hu}, W., \& {Kravtsov}, A.~V. 2003, \apj, 584, 702, \dodoi{10.1086/345846}

\bibitem[{Hunter(2007)}]{Hunter:2007}
Hunter, J.~D. 2007, Computing in Science \& Engineering, 9, 90, \dodoi{10.1109/MCSE.2007.55}

\bibitem[{{Johnson} {et~al.}(2015){Johnson}, {Chen}, \& {Mulchaey}}]{Johnson_2015}
{Johnson}, S.~D., {Chen}, H.-W., \& {Mulchaey}, J.~S. 2015, \mnras, 449, 3263, \dodoi{10.1093/mnras/stv553}

\bibitem[{{Kauffmann} {et~al.}(2003){Kauffmann}, {Heckman}, {White}, {Charlot}, {Tremonti}, {Brinchmann}, {Bruzual}, {Peng}, {Seibert}, {Bernardi}, {Blanton}, {Brinkmann}, {Castander}, {Cs{\'a}bai}, {Fukugita}, {Ivezic}, {Munn}, {Nichol}, {Padmanabhan}, {Thakar}, {Weinberg}, \& {York}}]{Kauffmann_2003}
{Kauffmann}, G., {Heckman}, T.~M., {White}, S. D.~M., {et~al.} 2003, \mnras, 341, 33, \dodoi{10.1046/j.1365-8711.2003.06291.x}

\bibitem[{{Kwak} \& {Shelton}(2010)}]{Kwak_shelton_2010}
{Kwak}, K., \& {Shelton}, R.~L. 2010, \apj, 719, 523, \dodoi{10.1088/0004-637X/719/1/523}

\bibitem[{{Lan}(2020)}]{Lan_2020}
{Lan}, T.-W. 2020, \apj, 897, 97, \dodoi{10.3847/1538-4357/ab989a}

\bibitem[{{Lehner} \& {Howk}(2011)}]{Lehner_howk_2011}
{Lehner}, N., \& {Howk}, J.~C. 2011, Science, 334, 955, \dodoi{10.1126/science.1209069}

\bibitem[{{Lehner} {et~al.}(2018){Lehner}, {Wotta}, {Howk}, {O'Meara}, {Oppenheimer}, \& {Cooksey}}]{Lehner_2018}
{Lehner}, N., {Wotta}, C.~B., {Howk}, J.~C., {et~al.} 2018, \apj, 866, 33, \dodoi{10.3847/1538-4357/aadd03}

\bibitem[{pandas~development team(2024)}]{pandas_2024}
pandas~development team, T. 2024, {pandas-dev/pandas: Pandas}, v2.2.1,  Zenodo, \dodoi{10.5281/zenodo.10697587}

\bibitem[{{Peeples} {et~al.}(2014){Peeples}, {Werk}, {Tumlinson}, {Oppenheimer}, {Prochaska}, {Katz}, \& {Weinberg}}]{Peeples_2014}
{Peeples}, M.~S., {Werk}, J.~K., {Tumlinson}, J., {et~al.} 2014, \apj, 786, 54, \dodoi{10.1088/0004-637X/786/1/54}

\bibitem[{{P{\'e}roux} \& {Howk}(2020)}]{peroux_howk_2020}
{P{\'e}roux}, C., \& {Howk}, J.~C. 2020, \araa, 58, 363, \dodoi{10.1146/annurev-astro-021820-120014}

\bibitem[{{Planck Collaboration} {et~al.}(2016){Planck Collaboration}, {Ade}, {Aghanim}, {Arnaud}, {Ashdown}, {Aumont}, {Baccigalupi}, {Banday}, {Barreiro}, {Bartlett}, {Bartolo}, {Battaner}, {Battye}, {Benabed}, {Beno{\^\i}t}, {Benoit-L{\'e}vy}, {Bernard}, {Bersanelli}, {Bielewicz}, {Bock}, {Bonaldi}, {Bonavera}, {Bond}, {Borrill}, {Bouchet}, {Boulanger}, {Bucher}, {Burigana}, {Butler}, {Calabrese}, {Cardoso}, {Catalano}, {Challinor}, {Chamballu}, {Chary}, {Chiang}, {Chluba}, {Christensen}, {Church}, {Clements}, {Colombi}, {Colombo}, {Combet}, {Coulais}, {Crill}, {Curto}, {Cuttaia}, {Danese}, {Davies}, {Davis}, {de Bernardis}, {de Rosa}, {de Zotti}, {Delabrouille}, {D{\'e}sert}, {Di Valentino}, {Dickinson}, {Diego}, {Dolag}, {Dole}, {Donzelli}, {Dor{\'e}}, {Douspis}, {Ducout}, {Dunkley}, {Dupac}, {Efstathiou}, {Elsner}, {En{\ss}lin}, {Eriksen}, {Farhang}, {Fergusson}, {Finelli}, {Forni}, {Frailis}, {Fraisse}, {Franceschi}, {Frejsel}, {Galeotta}, {Galli}, {Ganga}, {Gauthier}, {Gerbino}, {Ghosh}, {Giard},
  {Giraud-H{\'e}raud}, {Giusarma}, {Gjerl{\o}w}, {Gonz{\'a}lez-Nuevo}, {G{\'o}rski}, {Gratton}, {Gregorio}, {Gruppuso}, {Gudmundsson}, {Hamann}, {Hansen}, {Hanson}, {Harrison}, {Helou}, {Henrot-Versill{\'e}}, {Hern{\'a}ndez-Monteagudo}, {Herranz}, {Hildebrandt}, {Hivon}, {Hobson}, {Holmes}, {Hornstrup}, {Hovest}, {Huang}, {Huffenberger}, {Hurier}, {Jaffe}, {Jaffe}, {Jones}, {Juvela}, {Keih{\"a}nen}, {Keskitalo}, {Kisner}, {Kneissl}, {Knoche}, {Knox}, {Kunz}, {Kurki-Suonio}, {Lagache}, {L{\"a}hteenm{\"a}ki}, {Lamarre}, {Lasenby}, {Lattanzi}, {Lawrence}, {Leahy}, {Leonardi}, {Lesgourgues}, {Levrier}, {Lewis}, {Liguori}, {Lilje}, {Linden-V{\o}rnle}, {L{\'o}pez-Caniego}, {Lubin}, {Mac{\'\i}as-P{\'e}rez}, {Maggio}, {Maino}, {Mandolesi}, {Mangilli}, {Marchini}, {Maris}, {Martin}, {Martinelli}, {Mart{\'\i}nez-Gonz{\'a}lez}, {Masi}, {Matarrese}, {McGehee}, {Meinhold}, {Melchiorri}, {Melin}, {Mendes}, {Mennella}, {Migliaccio}, {Millea}, {Mitra}, {Miville-Desch{\^e}nes}, {Moneti}, {Montier}, {Morgante}, {Mortlock},
  {Moss}, {Munshi}, {Murphy}, {Naselsky}, {Nati}, {Natoli}, {Netterfield}, {N{\o}rgaard-Nielsen}, {Noviello}, {Novikov}, {Novikov}, {Oxborrow}, {Paci}, {Pagano}, {Pajot}, {Paladini}, {Paoletti}, {Partridge}, {Pasian}, {Patanchon}, {Pearson}, {Perdereau}, {Perotto}, {Perrotta}, {Pettorino}, {Piacentini}, {Piat}, {Pierpaoli}, {Pietrobon}, {Plaszczynski}, {Pointecouteau}, {Polenta}, {Popa}, {Pratt}, {Pr{\'e}zeau}, {Prunet}, {Puget}, {Rachen}, {Reach}, {Rebolo}, {Reinecke}, {Remazeilles}, {Renault}, {Renzi}, {Ristorcelli}, {Rocha}, {Rosset}, {Rossetti}, {Roudier}, {Rouill{\'e} d'Orfeuil}, {Rowan-Robinson}, {Rubi{\~n}o-Mart{\'\i}n}, {Rusholme}, {Said}, {Salvatelli}, {Salvati}, {Sandri}, {Santos}, {Savelainen}, {Savini}, {Scott}, {Seiffert}, {Serra}, {Shellard}, {Spencer}, {Spinelli}, {Stolyarov}, {Stompor}, {Sudiwala}, {Sunyaev}, {Sutton}, {Suur-Uski}, {Sygnet}, {Tauber}, {Terenzi}, {Toffolatti}, {Tomasi}, {Tristram}, {Trombetti}, {Tucci}, {Tuovinen}, {T{\"u}rler}, {Umana}, {Valenziano}, {Valiviita}, {Van Tent},
  {Vielva}, {Villa}, {Wade}, {Wandelt}, {Wehus}, {White}, {White}, {Wilkinson}, {Yvon}, {Zacchei}, \& {Zonca}}]{Plank2016}
{Planck Collaboration}, {Ade}, P.~A.~R., {Aghanim}, N., {et~al.} 2016, \aap, 594, A13, \dodoi{10.1051/0004-6361/201525830}

\bibitem[{{Prochaska} {et~al.}(2017){Prochaska}, {Werk}, {Worseck}, {Tripp}, {Tumlinson}, {Burchett}, {Fox}, {Fumagalli}, {Lehner}, {Peeples}, \& {Tejos}}]{prochaska_2017_metallicity}
{Prochaska}, J.~X., {Werk}, J.~K., {Worseck}, G., {et~al.} 2017, \apj, 837, 169, \dodoi{10.3847/1538-4357/aa6007}

\bibitem[{Prochaska {et~al.}(2017{\natexlab{a}})Prochaska, Tejos, Crighton, jnburchett, tiffanyhsyu, Tuo-Ji, marijana777, ktirimba, jhennawi, Cooke, O'Meara, \& Werk}]{prochaska_linetools_2017}
Prochaska, J.~X., Tejos, N., Crighton, N., {et~al.} 2017{\natexlab{a}}, {Linetools/Linetools: Third Minor Release}, v0.3,  Zenodo, \dodoi{10.5281/zenodo.1036773}

\bibitem[{Prochaska {et~al.}(2017{\natexlab{b}})Prochaska, Tejos, cwotta, jnburchett, Fumagalli, marijana777, O'Meara, Werk, Marc, mneeleman, \& kheegan}]{prochaska_pyigm_2017}
Prochaska, J.~X., Tejos, N., cwotta, {et~al.} 2017{\natexlab{b}}, {Pyigm/Pyigm: Initial release for publications}, v1.0,  Zenodo, \dodoi{10.5281/zenodo.1045479}

\bibitem[{{Rubin} {et~al.}(2014){Rubin}, {Prochaska}, {Koo}, {Phillips}, {Martin}, \& {Winstrom}}]{Rubin_2014}
{Rubin}, K. H.~R., {Prochaska}, J.~X., {Koo}, D.~C., {et~al.} 2014, \apj, 794, 156, \dodoi{10.1088/0004-637X/794/2/156}

\bibitem[{{Saintonge} \& {Catinella}(2022)}]{saintonge_catinella_2022}
{Saintonge}, A., \& {Catinella}, B. 2022, \araa, 60, 319, \dodoi{10.1146/annurev-astro-021022-043545}

\bibitem[{{Schiminovich} {et~al.}(2007){Schiminovich}, {Wyder}, {Martin}, {Johnson}, {Salim}, {Seibert}, {Treyer}, {Budav{\'a}ri}, {Hoopes}, {Zamojski}, {Barlow}, {Forster}, {Friedman}, {Morrissey}, {Neff}, {Small}, {Bianchi}, {Donas}, {Heckman}, {Lee}, {Madore}, {Milliard}, {Rich}, {Szalay}, {Welsh}, \& {Yi}}]{Schiminovich_2007}
{Schiminovich}, D., {Wyder}, T.~K., {Martin}, D.~C., {et~al.} 2007, \apjs, 173, 315, \dodoi{10.1086/524659}

\bibitem[{{Stocke} {et~al.}(2006){Stocke}, {Penton}, {Danforth}, {Shull}, {Tumlinson}, \& {McLin}}]{Stocke_2006}
{Stocke}, J.~T., {Penton}, S.~V., {Danforth}, C.~W., {et~al.} 2006, \apj, 641, 217, \dodoi{10.1086/500386}

\bibitem[{{Tchernyshyov} {et~al.}(2022){Tchernyshyov}, {Werk}, {Wilde}, {Prochaska}, {Tripp}, {Burchett}, {Bordoloi}, {Howk}, {Lehner}, {O'Meara}, {Tejos}, \& {Tumlinson}}]{Tchernyshyov2022}
{Tchernyshyov}, K., {Werk}, J.~K., {Wilde}, M.~C., {et~al.} 2022, arXiv e-prints, arXiv:2211.06436, \dodoi{10.48550/arXiv.2211.06436}

\bibitem[{{Tchernyshyov} {et~al.}(2023){Tchernyshyov}, {Werk}, {Wilde}, {Prochaska}, {Tripp}, {Burchett}, {Bordoloi}, {Howk}, {Lehner}, {O'Meara}, {Tejos}, \& {Tumlinson}}]{Tchernyshyov_2023}
---. 2023, \apj, 949, 41, \dodoi{10.3847/1538-4357/acc86a}

\bibitem[{{Tejos} {et~al.}(2016){Tejos}, {Prochaska}, {Crighton}, {Morris}, {Werk}, {Theuns}, {Padilla}, {Bielby}, \& {Finn}}]{Tejos_2016}
{Tejos}, N., {Prochaska}, J.~X., {Crighton}, N. H.~M., {et~al.} 2016, \mnras, 455, 2662, \dodoi{10.1093/mnras/stv2376}

\bibitem[{{Tumlinson} {et~al.}(2017){Tumlinson}, {Peeples}, \& {Werk}}]{tumlinson_2017}
{Tumlinson}, J., {Peeples}, M.~S., \& {Werk}, J.~K. 2017, \araa, 55, 389, \dodoi{10.1146/annurev-astro-091916-055240}

\bibitem[{{Tumlinson} {et~al.}(2011){Tumlinson}, {Thom}, {Werk}, {Prochaska}, {Tripp}, {Weinberg}, {Peeples}, {O'Meara}, {Oppenheimer}, {Meiring}, {Katz}, {Dav{\'e}}, {Ford}, \& {Sembach}}]{Tumlinson_2011}
{Tumlinson}, J., {Thom}, C., {Werk}, J.~K., {et~al.} 2011, Science, 334, 948, \dodoi{10.1126/science.1209840}

\bibitem[{{Tumlinson} {et~al.}(2013){Tumlinson}, {Thom}, {Werk}, {Prochaska}, {Tripp}, {Katz}, {Dav{\'e}}, {Oppenheimer}, {Meiring}, {Ford}, {O'Meara}, {Peeples}, {Sembach}, \& {Weinberg}}]{Tumlinson_2013}
---. 2013, \apj, 777, 59, \dodoi{10.1088/0004-637X/777/1/59}

\bibitem[{Virtanen {et~al.}(2020)Virtanen, Gommers, Oliphant, Haberland, Reddy, Cournapeau, Burovski, Peterson, Weckesser, Bright, {van der Walt}, Brett, Wilson, Millman, Mayorov, Nelson, Jones, Kern, Larson, Carey, Polat, Feng, Moore, {VanderPlas}, Laxalde, Perktold, Cimrman, Henriksen, Quintero, Harris, Archibald, Ribeiro, Pedregosa, {van Mulbregt}, \& {SciPy 1.0 Contributors}}]{2020SciPy-NMeth}
Virtanen, P., Gommers, R., Oliphant, T.~E., {et~al.} 2020, Nature Methods, 17, 261, \dodoi{10.1038/s41592-019-0686-2}

\bibitem[{{Werk} {et~al.}(2012){Werk}, {Prochaska}, {Thom}, {Tumlinson}, {Tripp}, {O'Meara}, \& {Meiring}}]{werk_2012}
{Werk}, J.~K., {Prochaska}, J.~X., {Thom}, C., {et~al.} 2012, \apjs, 198, 3, \dodoi{10.1088/0067-0049/198/1/3}

\bibitem[{{Werk} {et~al.}(2013){Werk}, {Prochaska}, {Thom}, {Tumlinson}, {Tripp}, {O'Meara}, \& {Peeples}}]{werk_2013}
---. 2013, \apjs, 204, 17, \dodoi{10.1088/0067-0049/204/2/17}

\bibitem[{{Werk} {et~al.}(2014){Werk}, {Prochaska}, {Tumlinson}, {Peeples}, {Tripp}, {Fox}, {Lehner}, {Thom}, {O'Meara}, {Ford}, {Bordoloi}, {Katz}, {Tejos}, {Oppenheimer}, {Dav{\'e}}, \& {Weinberg}}]{Werk_2014}
{Werk}, J.~K., {Prochaska}, J.~X., {Tumlinson}, J., {et~al.} 2014, \apj, 792, 8, \dodoi{10.1088/0004-637X/792/1/8}

\bibitem[{{Zahedy} {et~al.}(2019){Zahedy}, {Chen}, {Johnson}, {Pierce}, {Rauch}, {Huang}, {Weiner}, \& {Gauthier}}]{Zahedy_2019}
{Zahedy}, F.~S., {Chen}, H.-W., {Johnson}, S.~D., {et~al.} 2019, \mnras, 484, 2257, \dodoi{10.1093/mnras/sty3482}

\bibitem[{{Zheng} {et~al.}(2024){Zheng}, {Faerman}, {Oppenheimer}, {Putman}, {McQuinn}, {Kirby}, {Burchett}, {Telford}, {Werk}, \& {Kim}}]{Zheng_2024}
{Zheng}, Y., {Faerman}, Y., {Oppenheimer}, B.~D., {et~al.} 2024, \apj, 960, 55, \dodoi{10.3847/1538-4357/acfe6b}

\end{thebibliography}
\bibliographystyle{aasjournal}



\end{document}